\documentclass[12pt]{iopart}
\usepackage{iopams}

\expandafter\let\csname equation*\endcsname\relax
\expandafter\let\csname endequation*\endcsname\relax

\usepackage{amsmath}
\usepackage{amssymb}  

\usepackage{graphicx}

\usepackage{cite}
\usepackage{dsfont}
\usepackage[colorlinks, citecolor = blue, urlcolor = blue]{hyperref}
\usepackage{color}

\begin{document}

\title[On Landauer--B\"uttiker formalism from a quantum quench]{On Landauer--B\"uttiker formalism from a quantum quench}

\author{O Gamayun$^1$, Yu Zhuravlev$^2$ and N Iorgov$^{2,3}$}

\address{$^1$ Faculty of Physics, University of Warsaw, ul. Pasteura 5, 02-093 Warsaw, Poland}
\address{$^2$ Bogolyubov Institute for Theoretical Physics, 03143 Kyiv, Ukraine}
\address{$^3$ Kyiv Academic University, 03142 Kyiv, Ukraine}
\eads{\mailto{oleksandr.gamayun@fuw.edu.pl}, \mailto{ujpake@gmail.com}, \mailto{iorgov@bitp.kiev.ua}}

\vspace{10pt}
\begin{indented}
\item[]November 2022
\end{indented}

\begin{abstract}
We study transport in the free fermionic one-dimensional systems subjected to arbitrary local potentials. The bias needed for the transport is modeled 
by the initial highly non-equilibrium distribution where only half of the system is populated. Additionally to that, the local potential is also suddenly changed when the transport starts. For such a quench protocol we compute the Full Counting Statistics (FCS) of the number of particles in the initially empty part.
In the thermo\-dynamic limit, the FCS can be expressed via the Fredholm determinant with the kernel depending
on the scattering data and Jost solutions of the pre-quench and the post-quench potentials. 
We discuss the large-time asymptotic behavior of the obtained determinant and observe that if two or more bound states are present in the spectrum of the post-quench potential the information about the initial state manifests itself in the 
persistent oscillations of the FCS. On the contrary, when there are no bound states the asymptotic behavior of the FCS is determined solely by the scattering data of the post-quench potential,
which for the current (the first moment) is given by the Landauer--B\"uttiker formalism. 
The information about the initial state can be observed only in the transient dynamics.
\end{abstract}

\vspace{2pc}

\submitto{\JPA}
%\maketitle

\section{Introduction}

The Landauer--B\"uttiker formalism lies in the heart of mesoscopic physics \cite{Landauer_1957,doi:10.1080/14786437008238472,PhysRevLett.57.1761}. 
It directly allows one to express the conductance in terms of the transmission matrix, this way relating transport and quantum properties \cite{Landauer_1992,Imry_1999}. 
Historically, the substantiation of this formalism via linear response theory was connected with certain controversies ({\it cf} \cite{PhysRevLett.46.618,PhysRevB.23.6851} and \cite{PhysRevB.22.3519,PhysRevLett.47.972,PhysRevB.24.2978,PhysRevB.24.1151}). The original Landauer formulas proved to be sensitive to the proper formulation of the physical problem, in particular, to the proper definition of leads, electron reservoirs, and self-consistency of linear response (for review see \cite{Stone_1988}). 
The controversies were finally resolved by B\"uttiker in \cite{PhysRevLett.57.1761}, where the general formulas for multi-terminal mesoscopic conductance were obtained. 

Even though according to the elementary theory of tunneling the transmission probability is defined in a stationary setup there was a lot of attention 
related to the non-equilibrium approach to the transport \cite{Caroli_1971,PhysRevB.22.5887}. 
The powerful analytic approaches involving Keldysh Green's function techniques were developed in \cite{Stefanucci2004,PhysRevB.69.195318,KOHLER2005,Ridley2022}, along with the efficient numerical methods \cite{Gaury2016,PhysRevB.93.134506,Kloss2021}, which 
allow one not only to describe creation of the asymptotic currents and address their properties beyond the linear response regime but also explore behavior of the generic time-dependent quantum transport \cite{PhysRevB.66.205320,Moskalets2011,PhysRevB.103.L041405}. 

From the point of view of the one-dimensional integrable models the attention to similar problems was renewed in the context of the quantum quenches, which are specifically,
understood as the evolution of the isolated quantum system initialized in the highly non-equilibrium state created either via the rapid change of the Hamiltonian or containing macroscopic 
spatial inhomogeneities \cite{Calabrese_2007,Sotiriadis2008,Polkovnikov2011,Calabrese_2016,Eisert_2015}. The latter is more pertinent to the quantum transport setup and is dubbed as the partition approach \cite{Caroli_1971,PhysRevB.69.195318}. 
The large-time behavior of such systems can be described by the generalized hydrodynamics 
\cite{Bertini_2016,Castro_Alvaredo_2016}, which allows one to get analytic treatment of the non-equilibrium steady currents, 
describe anomalous diffusion, and address the correlation functions (for review see the special issue \cite{Bastianello_2022}.  

The transport in the translational invariant systems of free fermions and their spin analogs attracted a lot of attention due to the possibility of obtaining analytic answers for 
the average number of particles and its variance \cite{antal1999transport,Antal_2008,lancaster2010quantum,Viti_2016} (see also a numerical study in
\cite{PhysRevA.90.023624}). 
Other aspects of the evolution of the bipartite system were studied in \cite{Perfetto2017,Jin2021}. 
More delicate observables such as Loschmidt echo and Full Counting Statistics (FCS) were addressed in \cite{Viti_2016,St_phan_2017,PhysRevLett.110.060602,Sasamoto}, where the connection to  random matrix theory was performed 
and the FCS was expressed in terms of Fredholm determinants. 
Other connections of one-dimensional fermions at equilibrium in external potentials and random matrix theory are reviewed in \cite{Dean2019}.

The simplest case when translational invariance is broken by a local defect in many cases also allows for analytic treatment. 
Among others we would like to emphasize research that studies entropy evolution \cite{eisler2009entanglement,eisler2012on_entanglement,Dubail_2017}, 
transport properties within the interacting resonant level model \cite{Bransch_del_2010,PhysRevB.82.205414,Bidzhiev_2017,Bidzhiev_2019}, as well as non-integrable Ising chain \cite{PhysRevB.99.180302}. 
The inclusion of the defect in the generalized hydrodynamic approach was performed in \cite{Bertini2016}, the peculiarities of the thermalization via the defect were discussed in  
 \cite{10.21468/SciPostPhys.12.2.060}, and effects of the attractive local potential quench in \cite{Rossi2021}.
 Ref. \cite{10.21468/SciPostPhys.6.1.004} deals with the exact evaluation of the current and charge distribution for the bipartite scenario 
 when the left part of the system is prepared in the fully decorrelated state (infinite temperature) and is connected via the defect with the empty right part. 
 Further, this type of quench was considered for the hopping defect for the arbitrary initial distributions in \cite{Gamayun2020}, where FCS, Loschmidt echo, and the entanglement entropy were computed. 
 In \cite{Schehr2022} analytic answers for the particle and energy currents as well as the full density distribution were obtained for the continuous system with a delta impurity.
 
In this paper, we study the continuous bipartite system with an \textit{arbitrary} defect localized around the middle of the system. 
We consider a bipartite quench protocol, in which initially the ``right'' part of the system is empty and the ``left'' part is filled up to some energy with fermions subjected 
to the local short-range potential $V_0(x)$, or distributed according to some probability (to model, for instance, the thermal initial state). 
After that, the dynamics of the whole is governed by the Hamiltonian with the local potential $V(x)$, which may, in principle, be different from $V_0(x)$. We compute the FCS of the number of particles in the right part of the system. 
We derive an expression for FCS in the form of Fredholm determinant that is expressed via the Jost functions that correspond to the potentials $V$ and $V_0$. 
This is an exact expression in the thermodynamic limit that describes both the transient dynamics and the formation of the non-equilibrium steady-state. 
We argue that in the absence of the bound states in the potential $V(x)$, the leading terms in the FCS are defined via the transmission coefficient of the potential $V(x)$ and are given by the Levitov--Lesovik formula \cite{LL,Levitov_1996,Sch_nhammer_2007} (with logarithmic corrections for zero temperature states).   
If two or more bound states are present in the system they affect even the properties of the steady state by introducing persistent oscillations with a frequency equal to the difference of energies between the bound states. Moreover, the amplitude of these oscillations depends on the Jost functions of the potential $V_0(x)$, this way retaining the memory of the initial state. This phenomenon can be observed already on the level of the current, where even for the constant bias the persistent oscillations are present on top of the constant Landauer--B\"uttiker contribution.
Similar dependencies of the initial correlation in the case when bound states are present in the system were observed in \cite{Khosravi2008,PhysRevB.92.165403}.
This effect seems to be overlooked in the traditional approach (see for instance footnote 54 in \cite{PhysRevB.46.12485}).

The paper is organized as follows. In Section~\ref{sec2} we recall definitions of the scattering data, the Jost states and adopt notations for one-dimensional systems. 
In Section~\ref{quenchSec} we formulate the problem and present the main results. 
The outline of the derivation of the main results is presented in Sections \ref{HardWall} and 
\ref{secKernel}. In Section~\ref{HardWall} we describe a construction of the wave functions in the finite system and in Section~\ref{secKernel} we discuss how to obtain the kernel for the Fredholm determinant. Section~\ref{current} contains derivation of the Landauer--B\"uttiker expression for the current and its modification in the case when multiple bound states are present in the systems. A short summary and outlook are presented in Section~\ref{summ}. Appendices deal with some details of the derivations and contain scattering data for a few exemplary potentials.

\section{General properties of scattering}\label{sec2}

In this section we briefly remind some general notions of the one-dimensional scattering on the local potential $V(x)$. 
The eigenvalue problem satisfies the Schrodinger equation
\begin{equation}
    H_V\Psi= \left(-\frac{d^2}{dx^2} +V(x)\right)\Psi = E \Psi.
\end{equation}
The locality means that the potential vanishes fast enough as $|x|\to \infty$. For all practical purposes we assume that the potential is nonzero only in the finite domain 
$|x|<\xi$. This way, for $|x|>\xi$ the wave functions that correspond to the energy $E=k^2$ are the plane waves $e^{\pm i k x}$.
So for every real $k \neq 0$ there exists a two-dimensional space of solutions.  
The typical basis in this space can be conveniently described by the Jost states $\psi_k$, $\varphi_k$  defined by their asymptotic behavior, namely
\begin{equation}\label{eigenvalue}
\psi_k(x) = e^{-ikx} + o(1),\qquad x\to +\infty,
\end{equation}
\begin{equation}
\varphi_k(x)= e^{-ikx} + o(1),\qquad x\to -\infty.
\end{equation}
For a real potential these states are connected to their complex conjugated counterparts as $\psi_{-k}(x) = \bar{\psi}_k(x)$,
$\varphi_{-k}(x) = \bar{\varphi}_k(x)$.
If additionally the potential is symmetric $V(x) = V(-x)$, then  $\psi_k(-x)$ and $\varphi_k(-x)$ are still eigenfunctions. Considering the asymptotic behavior one can conclude that in this  case $   \psi_k(-x) = \bar{\varphi}_k(x)$. 
Using  \eref{eigenvalue} we see that the Jost solutions satisfy the following integral equations
\begin{equation}\label{psiint}
\psi_k(x) = e^{-ik x} - \int\limits_x^\infty \frac{\sin(k(x-y))}{k}V(y) \psi_k(y) dy,
\end{equation}
\begin{equation}\label{phiint}
\varphi_k(x) = e^{-ik x} + \int\limits^x_{-\infty} \frac{\sin(k(x-y))}{k}V(y) \varphi_k(y) dy.
\end{equation} 
As both Jost solutions form a basis they are connected by the linear transformation, the transfer matrix, 
\begin{equation}\label{transfer}
\left(
\begin{array}{c}
	\varphi_k(x) \\
	\bar{\varphi}_k(x)
\end{array}
\right) = 
\mathcal{T}(k)
\left(
\begin{array}{c}
	\psi_k(x) \\
	\bar{\psi}_k(x)
\end{array}
\right),\qquad 
\mathcal{T}(k) = \left(
\begin{array}{cc}
	a_k & b_k\\
	\bar{b}_k & \bar{a}_k
\end{array}
\right).
\end{equation}
Note that for a real potential $a_{-k}=\bar{a}_k$, $b_{-k}=\bar{b}_k$, while for a symmetric potential  $b_k$ is purely imaginary.

Considering the Wronskian of the eigenvalue problem \eref{eigenvalue} we conclude that the transfer matrix is unimodular
\begin{equation}\label{uni}
\det \mathcal{T}(k) =|a_k|^2-|b_k|^2=	1.
\end{equation}
The transfer matrix $\mathcal{T}$ can be repacked into the $S$-matrix \cite{Newton1982} as follows 
\begin{equation}
S = \frac{1}{a_k}\left(
\begin{array}{cc}
	-\bar{b}_k & 1\\
	1 & b_k
\end{array}
\right).
\end{equation}
The unimodularity condition \eref{uni} means the unitarity for S-matrix $SS^+ =1$. 
The transmission and the reflection coefficients are defined as the squared absolute values of the off-diagonal and diagonal components of the S-matrix, respectively,
\begin{equation}\label{tran}
    T(E) = \frac{1}{|a_k|^2},\qquad R(E) = \frac{|b_k|^2}{|a_k|^2}. 
\end{equation}
Here we present them as the functions of energy $E = k^2$. The unitarity \eref{uni} guarantees that $T(E) + R(E) = 1$. 

The coefficient $a_k$ can be analytically continued to the upper half plane where it might have zeroes that correspond to the bound states. They are purely imaginary $k=i\varkappa$ so the corresponding energy is negative $E = -\varkappa^2$. In fact the analytic properties allow one to present (see for instance \cite{Novikov})
\begin{equation}
    \label{a}
a_k = \prod\limits_{n=1}^{N} \frac{k-i\varkappa_n}{k+i\varkappa_n}
\exp\left(\frac{1}{2\pi i}\int\limits_{-\infty}^{\infty}\frac{\log (1+|b_q|^2)}{q-k-i0}dq\right).
\end{equation}
To describe the wave function of a bound state we can use either $\varphi_k(x)$ and $\bar{\psi}_k(x)$ as both these functions can be analytically continued to the upper half plane. In fact, it turns out that they are proportional $\varphi_{i\varkappa}(x) = b_\varkappa \bar{\psi}_{i\varkappa}$. Taking into account the definition of transfer matrix \eref{transfer} this relation is hardly surprising and $b_{\varkappa}$ can be considered as an analytic continuation of the $b_k$, however, contrary to $a_k$  such continuation is not always possible, and the coefficient $b_\varkappa$ should be considered as additional scattering data.  

Finally, let us comment on the normalization conditions of the continuous spectrum. Similar to \cite{Novikov} we conclude that 
\begin{equation}
    \int\limits_{-\infty}^\infty dx \varphi_k(x) \bar{\psi}_q(x) = a_q \delta(k-q).
\end{equation}
Therefore the Green's function $G(x,y,t)$ defined as a solution of the Schrodinger equation in $x$ variable with the initial condition $G(x,y,t=0) = \delta(x-y)$, can be presented as 
\begin{equation}\label{Gsimple}
    G(x,y,t) = \int_C\frac{dk}{2\pi} \frac{\varphi_k(x) \bar{\psi}_k(y)}{a_k} e^{-itE_k}.
\end{equation}
As for the continuum spectrum the contour $C$ goes along the real line. We notice however that the integrand can be analytically continued in the upper half plane. Moreover, in this form we can easily take into account also contributions from the bound states. To do so the contour $C$ should run above all positions of zeroes of $a_k$ in the upper half plane (see figure~\ref{FigContours} below). 
Below we re-derive this presentation using wave functions in the box (hard-wall boundary conditions), and demonstrate how to express full counting statistics via the scattering data and Jost solutions.

\section{Quench protocol} \label{quenchSec}

The scattering states introduced in the previous section describe an infinite system. To correctly formulate transport problem we consider 
open (hard-wall) boundary conditions placed at $x=\pm R$, perform computations at finite $R$, and send $R\to\infty$ in the end of the computation. 
At the initial moment of time only the left part of the system $x<0$ is filled. Meaning that the single particle  wave functions  $\Lambda_q$  are non-zero only in the interval $x\in [-R,0]$, more formally
\begin{equation}\label{eq1}
    - \frac{d^2\Lambda_q}{dx^2}+V_0(x)\Lambda_q = q^2\Lambda_q,\qquad\qquad \Lambda_q(0) = \Lambda_q(-R) = 0.
\end{equation}
The post-quench  wave functions satisfies
\begin{equation}\label{eq2}
    - \frac{d^2\chi_k}{dx^2}+V(x)\chi_k = k^2\chi_k,\qquad\qquad \chi_k(-R) = \chi_k(R) = 0.
\end{equation}
The initial $N$-particle state of the system  $|{\rm in}\rangle$ is given in a Fock space by an ordered set of momenta $q_1<q_2<\dots< q_N$. Formally, it can be presented as a wedge product
\begin{equation}\label{vac}
    |{\rm in}\rangle  =  \Lambda_{q_1}\bigwedge \Lambda_{q_2} \dots \bigwedge\Lambda_{q_N},
\end{equation}
which in the coordinate space corresponds to a single Slater determinant. The case of the statistical ensemble in the $N\to \infty$ limit can be described by taking the typical distribution of $q_i$.
To characterize  many body dynamics we consider full counting statistics (FCS). It can be written as
\begin{equation}\label{FCS2}
    \mathcal{F}(\lambda,t) = \langle {\rm in}| e^{itH} e^{\lambda N_R} e^{-itH} |{\rm in} \rangle = \langle {\rm in}| e^{\lambda \int\limits_0^t d\tau J(\tau)} |{\rm in} \rangle,
\end{equation}
where $N_R$ is number of particles in right part of the system and $J(\tau)$ is the current through the point $x=0$. 
Introducing the resolution of the unity, we can formally present FCS as a form factor series
\begin{equation}\label{ff1}
 \mathcal{F}(\lambda,t) = \sum_{\textbf{k},\textbf{p}}\langle {\rm in}| \textbf{k} \rangle \langle \textbf{k} |  e^{\lambda N_R} |\textbf{p}\rangle\langle \textbf{p}  |{\rm in} \rangle 
 e^{it(E_{\textbf{k}}-E_{\textbf{p}})}.
\end{equation}
Here $|\textbf{k}\rangle$ and $|\textbf{p}\rangle$ are many-body states of the form \eqref{vac}. 
Therefore the overlaps and the matrix elements are the determinants of the Cauchy type matrices.

Due to the free fermionic structure of the initial state \eref{vac} the FCS can be presented as 
\begin{equation}\label{Fdet}
    \mathcal{F}(\lambda,t) = \det X_{ab},
\end{equation}
with indices $a$ and $b$ corresponding to the momenta in the initial state $|{\rm in}\rangle$, and the matrix elements are
\begin{equation}
	X_{ab} =\delta_{ab}+ (e^\lambda-1)\sum_{k,p}  \frac{(\Lambda_a,\chi_k)(\chi_k, P_>\chi_p)(\chi_p,\Lambda_b)}{\sqrt{(\Lambda_a,\Lambda_a)}(\chi_k,\chi_k)(\chi_p,\chi_p)\sqrt{(\Lambda_b,\Lambda_b)}} e^{it(E_k-E_p)}.
	\label{Xab}
\end{equation}
Here $P_>$ is a projector on the right part of the system i.e. $x\in [0, R)$. 
This formula can be obtained from \eref{ff1} using some variant of the Cauchy--Binet formula (the product of determinants is the determinant of product of matrices).
Our goal is to present \eref{Fdet} in the thermodynamic limit as a Fredholm determinant of some trace-class operator. Namely, we present 
\begin{equation}\label{kk}
 X_{ab} =\delta_{ab}+ \frac{\pi}{R} K(q_a,q_b)+ o(1/R)   
\end{equation}
so that FCS
in the thermodynamic limit $R\to\infty$ transforms into a Fredholm determinant 
\begin{equation}\label{Ftd}
    \mathcal{F} (\lambda,t) = \det X \to \det \left(1 + \rho \hat{K}\right),
\end{equation}
where $\rho$ in the density of the initial state and the operator $\hat{K}$ acts on the integrable functions $L^2(\mathds{R})$ via the convolution with the kernel $K(q,q')$, namely 
\begin{equation}
    \hat{K}f(q) = \int K(q,q')f(q')dq'. 
\end{equation}
We compute this kernel in Section~\ref{secKernel}. It can be presented as 
\begin{equation}
   K(q,q') = K_0(q,q') + \delta K(q,q'),
\end{equation}
where 
\begin{equation}\label{K0}
    K_0(q,q') = \frac{e^\lambda-1}{\pi} \sigma(q,q')  \frac{\sin \frac{t(E_q -E_{q'})}{2}}{E_q -E_{q'}}
\end{equation}
with
\begin{equation}
    \sigma(q,q') = \frac{i |\Phi_q(0)||\Phi_{q'}(0)|}{ \Phi_q(0) \Phi_{q'}(0) \bar{a}_qa_{q'}} \left(
    \bar{\psi}_{q'}(0) \partial_x \psi_q(0) - \psi_q(0) \partial_x \bar{\psi}_{q'}(0). 
    \right)
\end{equation}
Here $\psi_k$ are Jost solutions defined by equation~\eqref{psiint} and by $\Phi_k(x)$ we denote the Jost solution equation~\eref{phiint} on the potential $V_0$. The expression for $\delta K$ can be found in  Section~\ref{secKernel}. It contains, in particular, contributions from the bound states if they are present in the spectrum of $V(x)$. 
We see that the kernels are expressed via the scattering data and the Jost solutions. 
The separation on $K_0$ and $\delta K$ is done to facilitate the large $t$ asymptotic analysis. 
Namely, in this limit $\delta K$ contains only oscillating terms, while formally $K_0$ tends to a delta function. For this reasoning we can heuristically argue that the leading contribution to the FCS will be given by $K_0$ and $\delta K$ will results in smooth prefactor for FCS. 
For a specific lattice system this effect was observed in \cite{Gamayun2020}. 
Moreover, since $\sigma(q,q')$ is a smooth function we can replace it with diagonal values $\sigma(q,q')\to \sigma(q,q)$.
Further, taking into account that the Wronskian $\bar{\psi}_{q}(x) \partial_x \psi_q(x) - \psi_q(x) \partial_x \bar{\psi}_{q}(x)$ does not depend on $x$, which can be checked by the immediate
differentiation. We evaluate it at $x\to-\infty$ and  arrive to the conclusion that $\sigma(q,q)=2q/|a_q|^2$. 
This allows us to transform the kernel to act on the energy space instead of momentum. This way, we obtain a Fredholm determinant of the generalized sine-kernel type 
\begin{equation}
    \mathcal{F}(\lambda,t) \approx \tilde{C}(\lambda,t) \det \left(1 + \frac{e^\lambda-1}{\pi}\rho(E)T(E)\frac{\sin \frac{t(E-E')}{2}}{E-E'} \right).
\end{equation}
Here we have written a kernel of the integral operator. The prefactor $ \tilde{C}(\lambda,t)$ appeared due to discarding $\delta K$. 
Notice that in this form all information about the Jost function disappears and only the transmission coefficient $T(E)$ for the post-quench potential remains. 
Large  $t$ asymptotic behavior of the Fredholm determinant can be easily found either by solving the corresponding Riemann--Hilbert problem \cite{Kitanine_2009,Slavnov_2010,Kozlowski_2011aa} or using the effective form factors 
\cite{GIZ,chernowitz2021dynamics,PhysRevB.105.085145}. For the smooth distribution $\rho(E)$ the result reads 
\begin{equation}
    \mathcal{F}(\lambda,t) \approx C(\lambda,t) \mathcal{F}_s(\lambda,t) 
\end{equation}
with 
\begin{equation} \label{fcsLL}
    \log \mathcal{F}_s(\lambda,t) = \frac{t}{2\pi }\int \log (1 + (e^\lambda-1) \rho(E) T(E)) dE \equiv i t \int \nu_\lambda(E) dE.
\end{equation}
\begin{figure}
    \centering
    \includegraphics[width=\linewidth]{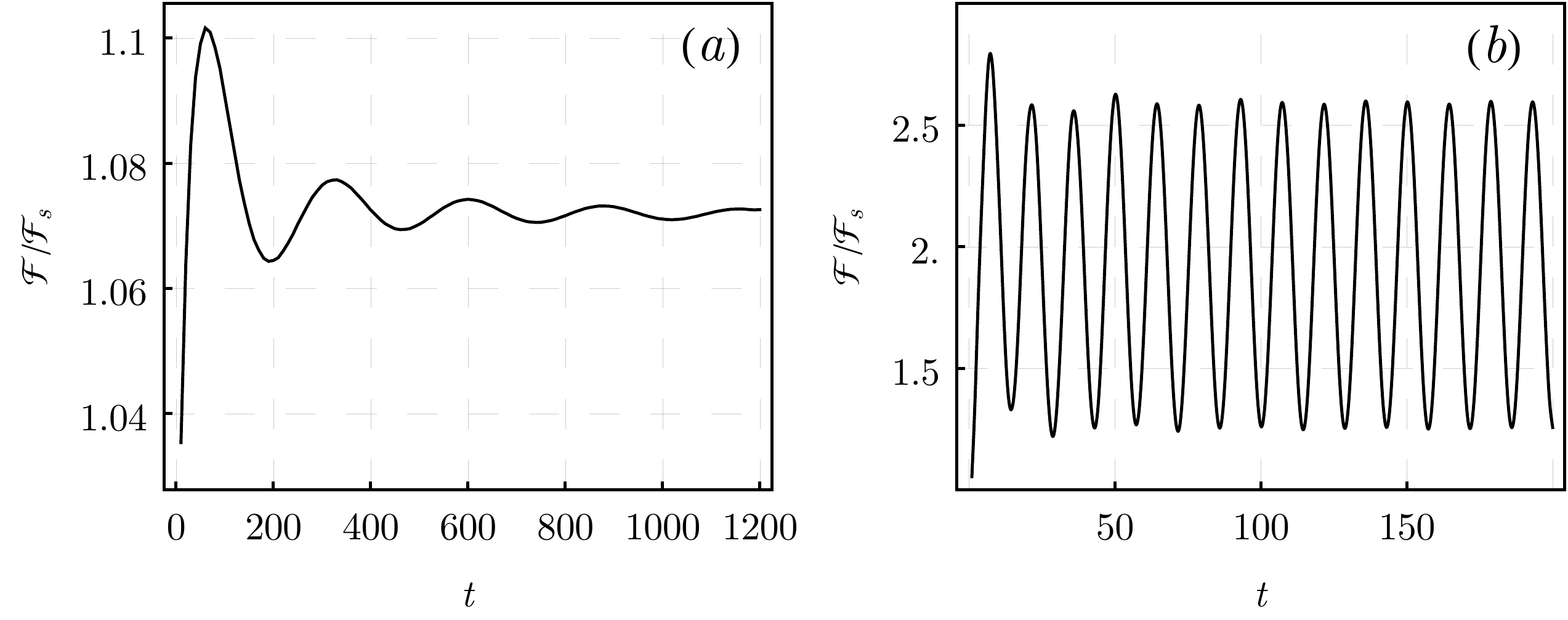}
    \caption{
    Ratio of the FCS $\mathcal{F}(\lambda,t)$ \eqref{Ftd} to the large $t$ asymptotic formula $\mathcal{F}_s(\lambda,t)$ given by \eref{fcsLL2}, the initial state is characterized by $k_F=1$, $E_F=k_F^2=1$, $\rho(E) = \theta(E_F-E)$:
    (a) delta barrier  $V(x)= g\delta(x)$, $g=-0.3$ (one bound state), $\lambda=0.3$;
    (b)  symmetric double delta barrier potential \eref{symdoubledelta} with $d=2.3$, $g=-1.3$ (two bound states), $\lambda=1.3$.
    }
    \label{FigFCS}
\end{figure}
The prefactor $C(\lambda,t)$ contains both $\tilde{C}(\lambda,t)$ and the constant prefactors from the asymptotic expression for the 
Fredholm determinant. 
When bound states are absent in the spectrum or there is only one bound state then we expect only decaying transient time dependence of $C(\lambda,t) \approx C(\lambda)$,
see figure~\ref{FigFCS}(a).
This way, in equation \eqref{fcsLL}, we recover predictions for the FCS also known as the Levitov--Lesovik formula 
\cite{LL,Levitov_1996,Sch_nhammer_2007}. The large deviation theory perspective on this formula can be found in \cite{RevModPhys.81.1665}, 
while the generalized hydrodynamic point of view is presented in \cite{Doyon2019}.
When the function $\rho(E)$ has sharp jumps, as it happens, for instance, at zero temperature $\rho(E) = \theta(E_F - E)$, or for the non-equilibrium 
setups  \cite{PhysRevB.81.085436,Gutman_2011}, then additionally to the smooth time dependence in $C(\lambda,t)$, we obtain also power law dependencies, with the corresponding exponents 
defined by the value of the function $\nu_\lambda(E)$ at the jump points. 
In particular, the modification of the vacuum case reads
\begin{equation}\label{fcsLL2}
 \log \mathcal{F}_s(\lambda,t) =- \left(\nu_\lambda(0)^2+ \nu_\lambda(E_F)^2\right) \log t +  
 \frac{t}{2\pi }\int\limits_0^{E_F} \log (1 + (e^\lambda-1)  T(E)) dE .
\end{equation}
Notice that $\nu_\lambda(0)=0$ for a generic barrier since $T(E=0)=0$. However for special potentials with $T(E=0)\ne 0$ (e.g. reflectionless potentials)   
$\nu_\lambda(0)\ne 0$ also gives a contribution to \eref{fcsLL2}.

Finally, when there are two or more bound states in the spectrum, then $C(\lambda,t)$ contains 
persistent oscillatory contributions with the frequency equal to the difference of energies of the bound states, see figure~\ref{FigFCS}(b).
Notice that after a few periods oscillations are described by one harmonic with a constant amplitude.
For a specific defect in a lattice model this was demonstrated in \cite{Gamayun2020}.

\subsection{Entanglement Entropy}

Let us also mention that one can relate the entanglement entropy $\mathcal{S}(t)$ 
obtained after tracing out the left part of the system to the FCS by a simple integral 
 \cite{Klich_2009,Klich_2009a,Song_2011,Song_2012}. 
 We express this relation in a simple and convenient form as
\begin{equation}\label{ee concise}
{\cal S}(t) = \frac{1}{4}  \int\limits_{-\infty}^\infty \frac{\log\mathcal{F}(\lambda,t) }{\sinh^2 (\lambda/2)}d \lambda,
\end{equation}
where the integral at $\lambda=0$  should be treated in the principal value sense.
Substituting instead of complete $\mathcal{F}$ its asymptotic expression $\mathcal{F}_s$ for instance for zero temperature case \eqref{fcsLL2}, we obtain  as $t\rightarrow \infty$
\begin{equation}\label{S asymptotic}
{\cal S}(t) \approx t  \int\limits_{0}^{E_F} \frac{dE}{2\pi} \Big(-T(E)  \log T(E) - R(E)\log R(E)\Big) - \frac{\log t }{4}  \int\limits_{-\infty}^\infty \frac{\nu_\lambda(0)^2+\nu_\lambda(E_F)^2}
{\sinh^2 (\lambda/2)}d \lambda ,
\end{equation}
Here $R(E) \equiv 1 -T(E)$. The linear in time part of this formula is generic for one-dimensional systems  \cite{Calabrese_2005E}, 
and in this case it has a form of classical Shannon entropy (see also \cite{eisler2012on_entanglement} and \cite{Bidzhiev_2017}), 
the suitable generalization to the interacting systems was obtained in \cite{Alba7947}. The logarithmic growth becomes important in the case of the absence of the defect, or for the
reflectionless potential, 
when the linear part disappears. The coefficient in front of the logarithm is compatible with predictions from conformal field theories \cite{calabrese2009entanglement,Peschel_2009,eisler2012on_entanglement}
\begin{equation}\label{S asymptotic epsilon=1}
{\cal S}(t) = \frac{c}{6} \log t+O(1) ,~~~ t\rightarrow \infty.
\end{equation}
In our case for $T(E) =1$ we get $c=2$ after computing the integral in the last line of \eqref{S asymptotic}. Notice that the coefficient in front of the logarithmic 
correction when the linear part is present can be non-universal similarly to \cite{eisler2012on_entanglement}.

\section{Hard-wall wave functions}
\label{HardWall} 
The key part in deriving explicit expression of kernels is an explicit presentation 
for the hard-wall wave functions \eqref{eq1}, \eqref{eq2} in terms of the Jost functions and scattering data. 
We start with $\chi_k$. Assuming that the range of the potential $\xi$ is much smaller than $R$, the wave function can be presented as 
\begin{equation}\label{chikk}
    \chi_k(x) = {\rm Im} \left[e^{ikR}\psi_k(x)\right],
\end{equation}
where $\psi_k$ is a Jost function that corresponds to the potential $V(x)$ (see \eref{psiint}). 
This way the condition $\chi_k(R) = 0 $ is satisfied automatically, while for the large negative $x$ the behavior reads 
\begin{equation}
    \chi_k(x) =  {\rm Im} \left[e^{ikR}(\bar{a}_k e^{-ikx} -b_k e^{ikx})\right].
\end{equation}
Here the scattering data corresponds to the potential $V(x)$. Demanding $\chi_k(-R)=0$ will provide us with the spectrum condition, that can be resolved as 
\begin{equation}\label{sp23}
	e^{2i k R} = \frac{i {\rm Im}\,b_k+\sqrt{1+({\rm Re}\,b_k)^2}}{\bar{a}_k} \equiv e^{-2i\delta(k)}.
\end{equation}
Here we have introduced the scattering phase $\delta(k)$. We have to take into account two possible solutions that correspond to two different branches of the square root. 
This way, in fact we have two different scattering phases. For both of them we have $\delta(k) = - \delta(-k)$,
meaning that if $k$ is a solution than $-k$ is solution as well, with the same energy $E_k = k^2$. 
However,  they describe the same state as is clearly seen from \eqref{chikk}. 
Therefore, we restrict ourselves to the positive $k$ solutions of \eref{sp23}. 

Let us also discuss the normalization of the wave function. 
To this end we notice that the $k$ derivative of the $\chi_k$ satisfies
\begin{equation}
\left(
-\partial_x^2+ V(x) - k^2
\right)\partial_k\chi_k = 2k \chi_k,\qquad 
\left(
-\partial_x^2+ V(x) - k^2
\right)\chi_k =0.
\end{equation}
So we can write 
\begin{multline}
2k (\chi_k,\chi_k) =  \int\limits_{-R}^{R} dx \left[
-\frac{d^2\partial_k\chi_k }{dx^2}\chi_k(x) + \partial_k\chi_k \frac{d^2\chi_k(x)}{dx^2}
\right] \\
=  \left[
-\frac{d\partial_k\chi_k }{dx}\chi_k(x) + \partial_k\chi_k \frac{d\chi_k(x)}{dx}
\right] \Big|_{-R}^{R}.\label{norm0}
\end{multline}
This allows us to present 
\begin{equation}\label{norm}
    (\chi_k,\chi_k) = ({\rm Re}\, b_k + \sqrt{1 + ({\rm Re}\, b_k)^2})\sqrt{1 + ({\rm Re}\, b_k)^2} (R + \delta'(k)).
\end{equation}
Here $\delta'(k)$ means the momentum derivative.
Similarly, we can describe the matrix elements $(\chi_k, P_>\chi_p)=\int\limits_{0}^R dx \chi_k(x) \chi_p(x) $ of the projector in \eref{Xab} as
\begin{multline}\label{chi2}
     (E_k-E_p)(\chi_k, P_>\chi_p) = \\
     =\int\limits_{0}^R dx \left(\left[\left(-\partial_x^2+ V(x) \right)\chi_k(x)\right] \chi_p(x) -
    \chi_k(x)\left(-\partial_x^2+ V(x) \right)\chi_p(x)\right)
      \\ =\int\limits_{0}^R dx \partial_x\left(
     -\chi_p(x) \partial_x\chi_k(x)+ \chi_k(x)\partial_x\chi_p(x)
     \right) = \chi_p(0) \partial_x\chi_k(0)-\chi_k(0) \partial_x\chi_p(0).
\end{multline}

To describe bound states that might be present in the system, one can argue that due to finite range of the potential the corresponding wave functions will be localized around $x=0$, and decay exponentially for large $x$. Therefore the boundary conditions are satisfied automatically with the exponential precision, and we may put
\begin{equation}
    \chi_k^{\rm bound} (x) \approx \varphi_{i\varkappa}(x),\qquad k = i\varkappa. 
\end{equation}
Its normalization can be found in a similar manner taking into account the identification  $\varphi_{i\varkappa}(x) = b_\varkappa \bar{\psi}_{i\varkappa}(x)$ discussed 
in Section~\ref{sec2}. 
Indeed, using the fact that at $x\to+\infty$, the leading term in the momentum in the wave function behaves as $a'_{i\varkappa } e^{\varkappa x}$, we obtain 
\begin{equation}\label{NormBound}
    (\varphi_{i\varkappa},\varphi_{i\varkappa}) = i a'_{i\varkappa} b_\varkappa.
 \end{equation}

Similarly we can find the pre-quench wave function $\Lambda_q$. In this case it is more convenient to use the Jost solution \eref{phiint} on the potential $V_0$, which we denote as 
$\Phi_q(x)$. In this notation we propose the following formula
\begin{equation}\label{lambda1}
    \Lambda_q(x) = {\rm Im}\frac{\Phi_q(x)}{\Phi_q(0)}.
\end{equation}
Notice that in this form the boundary condition $\Lambda_q(0)=0$ is satisfied automatically, while the condition $\Lambda_q(-R)=0$ defines spectrum
and the scattering phase $\eta(q)$
\begin{equation}\label{sp44}
    e^{2iqR} = \frac{\Phi_q(0)}{\bar{\Phi}_q(0)} \equiv e^{-2i\eta(q)}.
\end{equation}
Normalization now reads as 
\begin{equation}\label{LambdaOver}
    (\Lambda_q,\Lambda_q) =  \frac{R +\eta'(q)}{2|\Phi_q(0)|^2}.
\end{equation}
Finally, computation of the overlaps between pre- and post-quench wavefunctions in \eref{Xab}, can be avoided completely, and replaced by the corresponding 
overlaps with the Jost's functions. Namely, as it follows from  \eref{chi2} the time derivative of the $\eref{Xab}$ can be expressed via 
the (conjugated) time evolution of the wave function $\Lambda_q(y,t)$
defined as 
\begin{equation}\label{L0}
    \Lambda_q(y,t) \equiv \sum_k \frac{(\Lambda_q,\chi_k)\chi_k(y)}{(\chi_k,\chi_k)}e^{itE_k} = \int\limits_{-R}^0 dx \Lambda_q(x) G^*(x,y,t).
\end{equation}
Here we have used the following presentation of the Green's function 
\begin{equation}\label{Gstar}
    G^*(x,y,t) \equiv \sum\limits_k  \frac{\chi_k(x)\chi_k(y)}{(\chi_k,\chi_k)} e^{itE_k}.
\end{equation}
The summation is taken over all spectral points \eref{sp23}. We perform this summation explicitly in \ref{appG} with the genuine discrete degrees of freedom and take the thermodynamic limit only in the very end. The computation is straightforward but a bit tedious.
However, the obtained result can be easily explained heuristically. Namely, one can argue that in the thermodynamic limit instead of function \eref{Gstar}
one can use \eref{Gsimple}. 
This way,  we can find a presentation only with the Jost solutions introduced in the previous section
\begin{equation}
   \Lambda_q(y,t) =   \int_C \frac{dk}{2\pi} \frac{(\Lambda_q,\varphi_k)\bar{\psi}_k(y)}{a_k}e^{itE_k}.
\end{equation}
The integration path $C$ runs from $-\infty$ to $+\infty$ in the upper half plane above all positions of zeroes of $a_k$, see figure~\ref{FigContours}. 
The overlap $(\Lambda_q,\varphi_k)$ can be computed using the same trick as in  \eref{norm0} and  \eref{chi2}. 
Indeed, if we introduce function 
\begin{equation}\label{Xiqk}
    \Xi_{q,k} =\Lambda_q'(0)\varphi_k(0)- \int\limits_{-\infty}^0 dx \Lambda_q(x) (V_0(x) - V(x))\varphi_k(x),
\end{equation}
we can present 
\begin{equation}\label{overlll}
    (E_k -E_q) \int\limits_{-R}^0 dx \Lambda_q(x) \varphi_k(x) = \Xi_{q,k} - \Lambda_q'(-R)\varphi_k(-R).
\end{equation}
Here we have used that due to the finite range of the potentials the lower limit of the integration in \eref{Xiqk} can be either $-R$ or $-\infty$. 
Taking into account that for $k\in C$ the last term vanishes exponentially $\varphi_k(-R)\sim e^{ikR}$, we finally present 
\begin{equation}\label{L1}
     \Lambda_q(y,t) =   \int_C \frac{dk}{2\pi} \frac{\Xi_{q,k} \bar{\psi}_k(y)}{(k^2-q^2)a_k}e^{itE_k}.
\end{equation}
This is the final answer in the thermodynamic limit. 
Notice that $\Xi_{q,k}$ is a regular function and can be continued from the discrete spectrum to upper half plane of the variable $k$.
In the next section we will evaluate large-time asymptotic behavior of the kernel, which is mostly defined  by $\Xi_{q,-q}$. It can be 
computed from \eref{overlll} along with the asymptotic behavior $\Lambda_q'(-R)\sim -q e^{iqR}/\Phi_q(0)$ for large $R$ (see \eref{lambda1})  
\begin{equation}\label{Xiqmq}
	\Xi_{q,-q}=-\frac{q}{\Phi_q(0)}.
\end{equation}
This expression can be directly obtained from the definition \eref{Xiqk} already in the thermodynamic limit. We demonstrate it in \ref{pp}. 
The direct computation of $\Lambda_q(0,t)$ and its derivative in the finite system is given in \ref{appF}. 
%In the next section we use the presentation \eref{L1} to compute kernel $K(q,q')$ in \eref{kk} and \eref{Ftd}. 

\section{Kernel} \label{secKernel}

\begin{figure}
    \centering
    \includegraphics[width=0.8\linewidth]{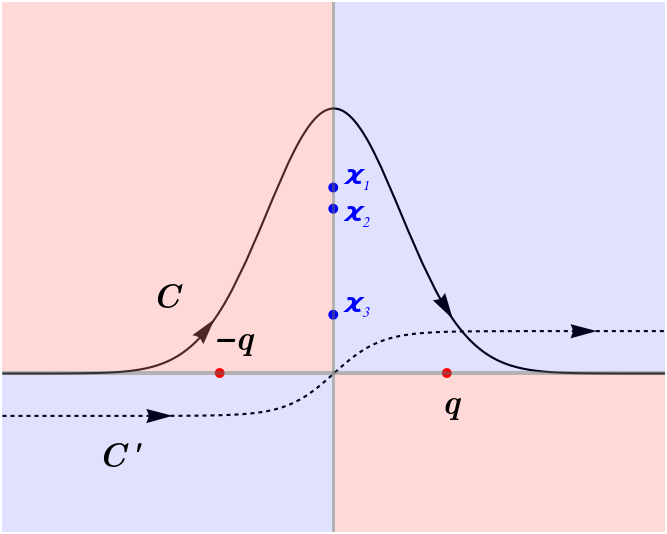}
    \caption{Integration contours $C$ and $C'$ in the complex plane of $k$ for the integral presentation of $f^{(\alpha)}_q $ given by \eref{fa2m}. 
    The contours $C$ are $C'$ are the initial and transformed contours of integration, respectively.
    Blue dots on the imaginary axis correspond to the bound states, red dots correspond to poles at $k=\pm q$
    in \eref{fa2m}.
    The shaded areas show the regions of exponential decaying (I, III quadrants, light blue)  and exponential growth (II, IV quadrants, pink)  of $\exp(it q^2)$ for $t\to+\infty$.
    }
    \label{FigContours}
\end{figure}

To compute the kernel $K(q,q')$ for the Fredholm determinant of the FCS \eref{Ftd}, we start by considering its time derivative. 
Using explicit presentation \eref{Xab} and \eref{chi2}, along with the definition \eref{L0}, we arrive at 
\begin{equation}\label{dK}
    \frac{dK(q,q')}{dt} = \frac{2i(e^\lambda-1)}{\pi}
    |\Phi_q(0)| \left( f^{(1)}_q(t) \bar{f}^{(0)}_{q'}(t) - f^{(0)}_q(t)\bar{f}^{(1)}_{q'}(t)\right)|\Phi_{q'}(0)|,
\end{equation}
where we have denoted 
\begin{equation}\label{fa2m}
f^{(\alpha)}_q (t) = \partial^\alpha_x \Lambda_q(x,t)\Big|_{x=0}=
\int\limits_C \frac{dk}{2\pi} 
	\frac{\Xi_{q,k}\partial_x^\alpha\bar\psi_k(0)}{a_k}
	\frac{e^{itk^2}}{k^2-q^2}, \qquad \alpha=0,1.    
\end{equation}
The contour $C$ runs as is shown in figure~\ref{FigContours}.
Using presentation \eref{L1} we can directly integrate  \eref{dK}. However, in order to easier assess the long-time asymptotic behavior we first identically transform $f^{(\alpha)}_q $ to highlight the most relevant terms as $t\to+\infty$. To do so we notice that the exponential $e^{itk^2}$ is decaying in the first and third  quadrants of complex plane of $k$ (see figure~\ref{FigContours}).
So we deform the contour $C$ into $C'$ by pulling it towards the real negative line and crossing it. 
By doing so we inevitably encircle all positions of the bound states and the pole $k=-q$. 
The obtained deformation reads 
\begin{equation}\label{faCp}
	f^{(\alpha)}_q(t) = \frac{i\Xi_{q,-q}\partial_x^\alpha\bar\psi_{-q}(0)}{a_{-q}} 	\frac{e^{itq^2}}{2q}+   \sum_{n=1}^{N^\mathrm{b}} 
	\frac{i \Xi_{q,i\varkappa_n}\partial_x^\alpha\bar\psi_{i\varkappa_n}(0)}{a'_{i\varkappa_n}}
	\frac{e^{-it\varkappa_n^2}}{\varkappa_n^2+q^2}
		+
	\int\limits_{C'} \frac{dk}{2\pi} 
	\frac{\Xi_{q,k}\partial_x^\alpha\bar\psi_k(0)}{a_k}
	\frac{e^{itk^2}}{k^2-q^2}.
\end{equation}
The ``leading'' coefficient $\Xi_{q,-q}$ was computed in \eqref{Xiqmq}. 
Further we use the symmetry $k\to -k$ to fold the full contour $C'$ and consider integration only with $\mathrm{Re}\,k>0$, namely
\begin{equation}\label{faRe}
	f^{(\alpha)}_q(t) =  \sum_{n=1}^{N^\mathrm{b}} 
	 B_{n,q}^{(\alpha)} e^{-it\varkappa_n^2}
		+F^{(\alpha)}_q e^{itq^2} + 
	\int\limits_{0}^{\infty} \frac{dk}{\pi} 
	\Omega^{(\alpha)}_{q,k}
	\frac{e^{itk^2}}{(k+i0)^2-q^2},
\end{equation}
\begin{equation}\label{BFOm}
    B_{n,q}^{(\alpha)}=
    \frac{i \Xi_{q,i\varkappa_n}\partial_x^\alpha\bar \psi_{i\varkappa_n}(0)}
    {a'_{i\varkappa_n}(\varkappa_n^2+q^2)},
    \qquad
    F_q^{(\alpha)}=-i \frac{\partial_x^\alpha\psi_{q}(0)}{2\Phi_q(0)a_{-q}},
    \qquad
    \Omega^{(\alpha)}_{q,k}=
    \mathrm{Re}\,
	\frac{\Xi_{q,k}\partial_x^\alpha\bar\psi_k(0)}{a_k}.
\end{equation}
Such form of $f^{(\alpha)}_q(t)$ is convenient for large $t$ asymptotic analysis. The 
first two terms give persistent oscillations, while the integral in \eref{faRe} is decaying as a power law in $t$ for large $t$. 
This can be deduced from the stationary phase method considering a saddle point at $k=0$. The corresponding exponent of the power law decay depends on the behavior of $\Omega^{(\alpha)}_{q,k}$ at $k=0$. In the case of generic potentials, $a_k$ has a first order pole at $k=0$  while 
$\Xi_{q,k}$ and $\partial_x^\alpha\psi_k(0)$ are regular at $k=0$. 
Therefore $\Omega^{(\alpha)}_{q,k}$ has at least first order zero at $k=0$,
which implies the entire integral to be estimated as $O(t^{-1})$.
For some special potentials (for example reflectionless potentials), $a_k$ is regular at $k=0$. For such potentials the integral decays
as $t^{-1/2}$
\begin{equation}\label{intreg}
    \int\limits_{0}^{\infty} \frac{dk}{\pi} 
	\Omega^{(\alpha)}_{q,k}
	\frac{e^{itk^2}}{(k+i0)^2-q^2}=\frac{I^{(\alpha)}_{q}}{\sqrt{t}}+ 	O(t^{-1}),
\end{equation}
\begin{equation}\label{Iq}
I^{(\alpha)}_{q}=- \frac{\sqrt{\pi} e^{i\pi/4} \Xi_{q,0}\partial_x^\alpha\psi_0(0)}{2 a_0 q^2}.
\end{equation}
To compute the kernel we substitute $f^{(\alpha)}(t)$ in the form \eref{faRe} into~\eref{dK} and integrate over $t$.
Additionally, we perform conjugation with diagonal matrices  
\begin{equation}
	K(q,q') \to K(q,q')e^{-it(E_q -E_{q'})/2}.
\end{equation}
This operation does not change the determinant, so for the transformed kernel we obtain
\begin{equation}\label{Kqqp}
	K(q,q') =K_0(q,q')+ \delta K (q,q').
\end{equation}
Here $K_0(q,q')$ is given by 
\begin{equation}
	K_0(q,q') = \frac{4i(e^\lambda-1)}{\pi} |\Phi_q(0)| (F_q^{(1)} \bar F_{q'}^{(0)}-F_q^{(0)} \bar F_{q'}^{(1)})|\Phi_{q'}(0)|
	\frac{\sin t(E_q - E_{q'})/2}{E_q - E_{q'}}.
\end{equation}
Using definition \eqref{BFOm} it can be equivalently presented as \eqref{K0}. 
The rest of the 
kernel can be presented as 
\begin{equation}
	\delta K(q,q')=	\frac{2i(e^\lambda-1)}{\pi}
	|\Phi_q(0)| \left( M_{qq'}(t)- \bar{M}_{q'q}(t) \right)|\Phi_{q'}(0)|
\end{equation}
with
\begin{equation}
	M_{qq'}(t) = e^{-it(E_q -E_{q'})/2}\sum\limits_{i=1}^4 \left[K^{(i)}(q,q',t)-K^{(i)}(q,q',0)\right].
\end{equation}
Here different kernels have different physical meaning. The kernel $K^{(1)}$ is responsible for the contribution of the 
bound states only. It is given by 
\begin{equation}
	K^{(1)}(q,q',t)= \sum_{m<n}^{N^\mathrm{b}} (B_{mq}^{(1)}B_{nq'}^{(0)}-B_{mq}^{(0)} B_{nq'}^{(1)})
	\frac{e^{it(\varkappa_n^2-\varkappa_m^2)}}{i(\varkappa_n^2-\varkappa_m^2)} .
\end{equation}
The kernel $K^{(2)}$ is responsible for contribution of the continuous spectrum only
\begin{multline}
	K^{(2)}(q,q',t)= \int_0^\infty \frac{dk}{\pi} \frac{e^{it(E_k-E_{q'})}}{i(E^+_k-E_{q'})} 
	\frac{ \Omega_{qk}^{(1)}\bar F_{q'}^{(0)}- \Omega_{qk}^{(0)}\bar F_{q'}^{(1)}}{E^+_k-E_q} \\
	+\frac12 \int\limits_0^\infty \frac{dk}{\pi}\int\limits_0^\infty \frac{dp}{\pi} \frac{e^{it(E_k-E_p)}}{i(E^+_k-E^-_p)}
	\frac{\Omega_{qk}^{(1)} \Omega_{q'p}^{(0)}-\Omega_{qk}^{(0)} \Omega_{q'p}^{(1)}}
	{(E^+_k-E_q)(E^-_p-E_{q'})},
\end{multline}
here $E_k=k^2$ and $E^\pm_k=(k\pm i0)^2$. 
Finally the kernels $K^{(3)}$ and $K^{(4)}$ give the mixed contribution from the bound states and the continuous spectrum 
\begin{equation}
 K^{(3)}(q,q',t)=\sum_{n=1}^{N^\mathrm{b}} 
	\int_0^\infty \frac{dk}{\pi} \frac{e^{it(E_k+\varkappa_n^2)}}{i(E^+_k+\varkappa_n^2)} 
	\frac{ \Omega_{qk}^{(1)} B_{nq'}^{(0)}- \Omega_{qk}^{(0)} B_{nq'}^{(1)}}{E^+_k-E_q},
\end{equation}
\begin{equation}
		K^{(4)}(q,q',t)=\sum_{n=1}^{N^\mathrm{b}} (B_{nq'}^{(0)} F_{q}^{(1)}-B_{nq'}^{(1)} F_{q}^{(0)})
	\frac{e^{it(\varkappa_n^2+E_{q})}}{i(E_{q}+\varkappa_n^2)}.
\end{equation}
Integrals in $K^{(2)}$ and $K^{(3)}$ decay for large $t$ because of averaging of rapid oscillations as in the integral \eqref{faRe}. 
Special care has to be taken for the reflectionless potentials. At the first glance, in this case relations \eref{intreg},
\eref{Iq} might produce a logarithmic growth for large $t$ in the double integral in $K^{(2)}$.
This growth is, however, absent because of the relation 
\begin{equation}\label{resreg}
	I^{(1)}_{q}\bar I^{(0)}_{q'}-I^{(0)}_{q}\bar I^{(1)}_{q'}=0.
\end{equation}
There are also potential singularities for small $q\lesssim t^{-1/2}$ and a bit different asymptotic analysis of  \eref{intreg} is needed. Indeed, \eref{Iq} shows a singular behavior for small $q$, which in fact is not there, since in  the asymptotic analysis of \eref{intreg} we have assumed  that a pole at $k=q$ is far from the stationary point $k=0$. We performed such analysis for the current and showed that the contribution of small $q$ gives only the subleading contributions. 

Apart from the decaying terms, $\delta K$ contains also time-independent terms  $K^{(i)}(t=0)$, highly oscillating terms like $K^{(4)}(t)$, 
and terms that oscillate with the frequencies given by the energies of the bound states $K^{(1)}(t)$. The latter comes in the form of the finite rank operators, 
and can appear in the final expression of the determinant only linearly. 
As we have discussed in Section~\ref{quenchSec} we expect that 
the contribution of the kernel $\delta K$ to the asymptotic analysis of the Fredholm operator $\det (1 + \hat{K})$ 
enters only as a smooth overall prefactor, which has non-vanishing time dependence only if there are two or more bound states in the spectrum.

\subsection{FCS for perfect lead attachment}
\label{perf}

There are well-developed methods for asymptotic analysis of the Fredholm determinants of the so-called integrable kernels \cite{Deift_1997,Bogoliubov1997}. 
As we have shown above for generic potentials $V_0(x)$ and $V(x)$ the kernel for FCS $K(q,q')$ is not an integrable one. 

In this subsection we consider a special case of quench setup when the obtained kernel is integrable even for finite times. 
We call this situation 	the \textit{perfect lead attachment} because it corresponds to the scenario when $V_0(x) = V(x)$ for $x<0$.

In this case due to the integral presentation \eref{phiint} the corresponding Jost functions coincide for negative $x$: $\varphi_q(x) =\Phi_q(x)$ for $x \le 0$. From  presentation \eref{Xiqk} we observe the factorization
\begin{equation}\label{XiqkPLA}
	\Xi_{q,k} = \Lambda'_q(0) \varphi_k(0),
\end{equation}
which imply a similar factorization ${f}^{(\alpha)}_q(t)= \Lambda'_q(0) g^{(\alpha)}_q(t)$ for  ${f}^{(\alpha)}_q(t)$ given  by \eref{fa2m},
where 
\begin{equation}\label{fa22}
	g^{(\alpha)}_q(t) =  \int\limits_C \frac{dk}{2\pi} \omega_k^{(\alpha)}
		\frac{e^{itk^2}}{k^2-q^2} ,\qquad
		\omega_k^{(\alpha)} \equiv \frac {\varphi_{k}(0)\partial_x^\alpha\bar\psi_k(0)}{a_k}.
\end{equation}
Comparing \eref{XiqkPLA} at $k=-q$ with \eref{Xiqmq} we conclude that 
$\Lambda'_q(0)= -q/|\varphi_q(0)|^2 $. 
Therefore now \eref{dK} reads
\begin{equation}\label{dKqqp}
 \frac{dK(q,q')}{dt} = \frac{2i(e^\lambda-1)qq'}{\pi |\varphi_q(0)| |\varphi_{q'}(0)|}
\left( g^{(1)}_q(t) \bar{g}^{(0)}_{q'}(t) - g^{(0)}_q(t)\bar{g}^{(1)}_{q'}(t)\right).
\end{equation}
Integrating in $t$ we can present the kernel $K(q,q')$ in the integrable form
\begin{equation}\label{Kqqpl}
	K(q,q') = \frac{2(e^\lambda-1)qq'}{\pi|\varphi_q(0)| |\varphi_{q'}(0)|} \frac{g^{(1)}_q(t) \bar{g}^{(0)}_{q'}(t) - g^{(0)}_q(t)\bar{g}^{(1)}_{q'}(t)+
	\bar{D}_q(t)-D_{q'}(t)}{E_q - E_{q'}},
\end{equation}
where 
\begin{equation}
	D_{q}(t) =i \int\limits_0^t d\tau \int\limits_C \frac{dk}{2\pi} e^{i\tau k^2}  
	\left[ \omega_k^{(1)}\bar{g}^{(0)}_{q}(\tau)-\omega_k^{(0)}\bar{g}^{(1)}_{q}(\tau) \right] .
\end{equation}
To check correctness of \eref{Kqqpl} we need to compare its derivative in $t$ with \eref{dKqqp} using 
\begin{equation}
	\frac{d}{dt}g^{(\alpha)}_q(t)  =iq^2 g^{(\alpha)}_q(t)+ i \int\limits_C \frac{dk}{2\pi} \omega_k^{(\alpha)}e^{i t k^2}.
\end{equation}
Also we have to check that $K(q,q')=0$ at $t=0$. This is ensured due to the property $g^{(0)}_q(0)=0$, which follows from analyticity of $\omega_k^{(0)}$ in the upper half-plane of $k$.
The integrable form of kernel $K(q,q')$ allows one to replace evaluation of the Fredholm determinants by a solution of the Riemann--Hilbert problem \cite{Deift_1997,Bogoliubov1997}.
This approach is especially useful for the asymptotic analysis at large time $t\to + \infty$.
In this case, however, if we follow the standard procedure outlined in \cite{Bogoliubov1997}, the corresponding jump matrix will have size $4\times4$.
Therefore, we postpone full analysis to a separate publication. 

The asymptotic behavior of $g^{(\alpha)}_q(t)$ can be found similarly to \eqref{faRe}, where one can neglect the last integral. 
To find the large-time asymptotic behavior of ${D}_{q}(t)$
we present it identically as
\begin{multline}
	{D}_{q}(t) = \int\limits_{C} \frac{dk}{2\pi} \int\limits_{C^*} \frac{dp}{2\pi}
	\frac{e^{it(k^2-p^2)}-1}{k^2-p^2} \frac{\bar{\omega}_p^{(0)}\omega_k^{(1)}-\bar{\omega}_p^{(1)}\omega_k^{(0)}}{p^2-q^2} \\ \approx 
	-\int\limits_{C} \frac{dk}{2\pi} \int\limits_{C^*} \frac{dp}{2\pi}
	\frac{1}{k^2-p^2+i0} \frac{\bar{\omega}_p^{(0)}\omega_k^{(1)}-\bar{\omega}_p^{(1)}\omega_k^{(0)}}{p^2-q^2}.
\end{multline}
Here $C^*$ is a contour conjugated to $C$. 
Moreover, for the symmetric potential function $g^{(1)}_q(t)$ simplifies significantly and the integral can be dropped even for finite times, namely, we can present 
\begin{equation}
	g^{(1)}_q(t) = \frac{e^{itq^2}}{2 \bar{a}_q}.
\end{equation}
Here we used that for arbitrary even potential $V(-x)=V(x)$, the Jost solutions are related as $\psi_{-k}(x)=\varphi_k(-x)$,
which leads to 
\begin{equation} \label{sym}
\omega^{(1)}_k = 	\frac{\varphi_k(0)\partial_x \psi_{-k}(0)}{a_k} = ik.
\end{equation}
Indeed taking into account that the Wronskian $\varphi_k(x) \partial_x \psi_{-k}(x) - \psi_{-k}(x) \partial_x \varphi_k(x)$ does not depend on $x$ and calculating it at $x\to -\infty $ and $x=0$
we obtain the relation \eqref{sym}. 
Thus, the integral in \eqref{faRe} vanishes identically, since it depend only on the real part of \eqref{sym}. 
Further the bound state contribution vanishes because  the wave-functions are either odd or even, meaning that either the value at zero or the value of the derivative at zero vanishes leading to $\varphi_{i\varkappa_n}(0)\partial_x \bar{\psi}_{i\varkappa_n}(0)=0$.

\section{The current}\label{current}

Let us also discuss the full current $J(t)$ of the particles flowing through the middle $x=0$ to the right part of the system. 
It can be evaluated from the FCS \eref{Ftd} as follows 
\begin{equation}\label{J}
    J(t) =\frac{d}{dt} \frac{d\mathcal{F}(\lambda,t)}{d\lambda}\Big|_{\lambda=0}  
    = \mathrm{Tr}\,\left(\rho \frac{d}{dt} \frac{d\hat K}{d\lambda}\Big|_{\lambda=0}\right) =
    -\int_0^\infty dq \rho(q) 
    \frac{4|\Phi_q(0)|^2 }{\pi}
      {\rm Im}\,  f^{(1)}_q(t) \bar{f}^{(0)}_{q}(t) ,
\end{equation}
where at the last step we used explicit presentation \eref{dK} to compute the trace.
As we discuss in Section \ref{secKernel}, the integral in \eref{faRe} may be dropped for the calculation of current for large $t$ since it 
vanishes as $t\to \infty$, and we can approximate
\begin{equation}\label{faap}
	f^{(\alpha)}_q(t) \approx F^{(\alpha)}_q e^{itq^2} + \sum_{n=1}^{N^\mathrm{b}} 
	 B_{n,q}^{(\alpha)} e^{-it\varkappa_n^2}.
\end{equation}
Substituting this expression into \eref{J} we obtain three type of contributions to the current
\begin{equation}
    J(t)\approx J_\mathrm{LB} + J^\mathrm{b} + \delta J,
\end{equation}
where $J_\mathrm{LB}$ comes from the first term in \eref{faap},
$J^\mathrm{b}$ comes from the terms that involve the bound states only and
$\delta J$ described the mix of the first term with the bound states.

To calculate  $J_\mathrm{LB}$ we use ${\rm Im}\, \psi_q'(0)\bar{\psi}_q(0) = -q$
and \eref{tran}
\begin{equation}
    J_\mathrm{LB}=\int \limits_{0}^\infty \frac{dq }{\pi} \frac{q\rho(q)}{|a_q|^2} = 
    \int \frac{dE}{2\pi} \rho(E) T(E) .
\end{equation}
It is well-known Landauer--B\"uttiker formula for the current. 

The contribution of bound states to the current is
\begin{equation}
    J^\mathrm{b}=\sum_{m<n}
    A_{mn} \sin t(\varkappa_m^2-\varkappa_n^2),
\end{equation}
where 
\begin{equation}\label{Amn}
    A_{mn} =\frac{  4 \left(\bar\psi_{i\varkappa_n}'(0)\bar\psi_{i\varkappa_m}(0)-\bar\psi_{i\varkappa_m}'(0)\bar\psi_{i\varkappa_n}(0)\right) }{a'_{i\varkappa_m}a'_{i\varkappa_n}}
    \int_0^\infty \frac{dq}{\pi} \rho(q) 
    |\Phi_q(0)|^2 
    \frac{ \Xi_{q,i\varkappa_m}\Xi_{q,i\varkappa_n}}
    {(\varkappa_m^2+q^2)(\varkappa_n^2+q^2)}    .
\end{equation}
For the symmetric potential $V(x)$, the bound states are either even functions with $\bar\psi_{i\varkappa_n}'(0)=0$ or odd functions with  
$\bar\psi_{i\varkappa_n}(0)=0$. Therefore, in this case, a nontrivial contribution to the current may arise only from pairs of odd-even states. Furthermore, in the case of perfect lead attachment, $V(x)=V_0(x)$, we have $\Xi_{q,i\varkappa_n}=0$ for odd bound states 
$\bar\psi_{i\varkappa_n}(x)$
and therefore there is no contribution at all to the current from bound states in the case of perfect lead attachment with an even potential. 

The integral in $q$ for $\delta J$ can be estimated by the
contribution at $q=0$ by the method of stationary phase and it can be shown that   $\delta J$ 
decays for large $t$ at least as $t^{-1/2}$ and therefore does not give a leading contribution to the current.

Finally we arrive to the following expression for the large-time asymptotic current 
\begin{equation}\label{Jtot}
	J(t)\approx \int \frac{dE}{2\pi} \rho(E) T(E) +
	\sum_{m<n}
	A_{mn} \sin t(\varkappa_m^2-\varkappa_n^2).
\end{equation}
We see that in addition to the constant  Landauer--B\"uttiker current 
(the first term), there are also oscillating terms connected with the presence of the multiple bound states.
\begin{figure}
    \centering
    \includegraphics[width=\linewidth]{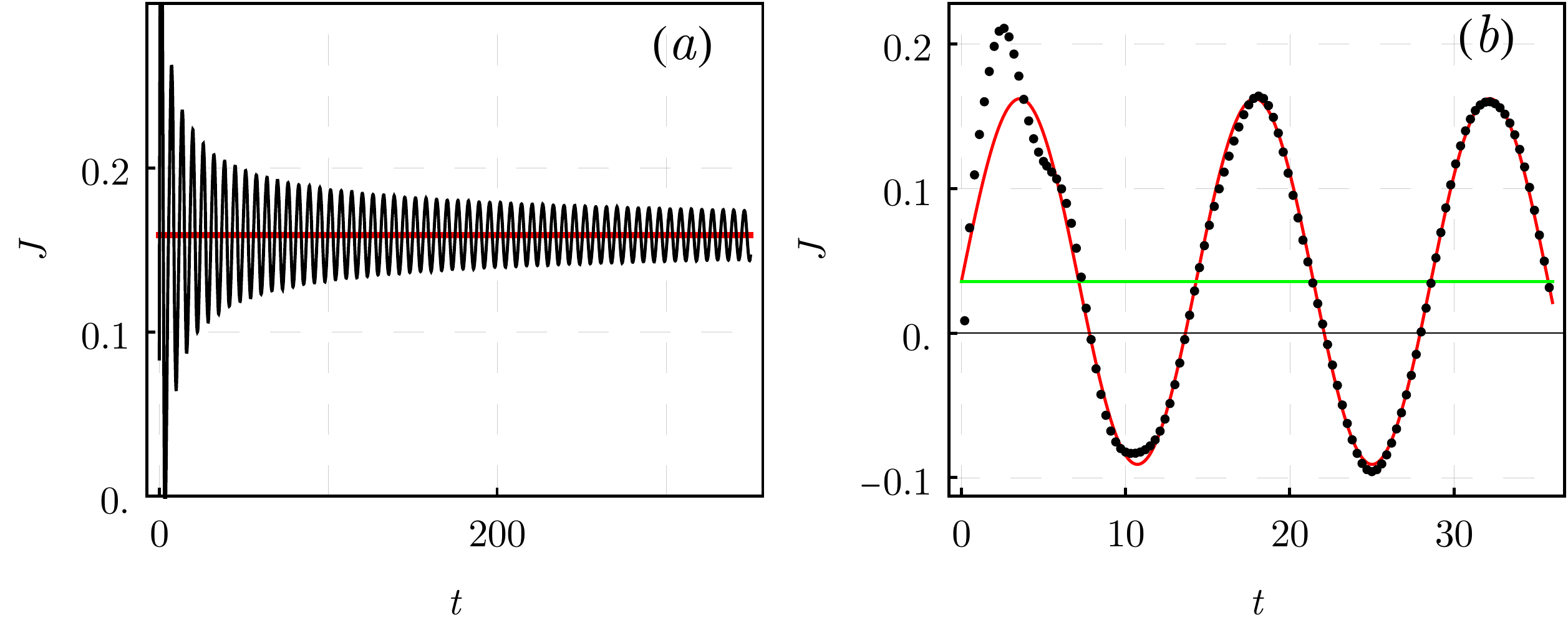}
    \caption{
    Current through the point $x=0$ and its asymptotic behavior, the initial state is characterized by $k_F=1$, $E_F=k_F^2=1$, $\rho(E) = \theta(E_F-E)$:
    (a) the reflectionless potential   $V(x)= - 2/\cosh^2 x$ (one bound state); the current (black) is oscillating with an amplitude decaying as $\sim t^{-1/2}$ around Landauer--B\"uttiker constant current $J_{LB} = E_F/(2\pi)$ (red).
    (b)  symmetric double delta barrier potential \eref{symdoubledelta} with $d=2.3$, $g=-1.3$ (two bound states); 
    the current (black dots) has asymptotic oscillating behavior \eref{Jtot2} with fixed amplitude (red curve) around Landauer--B\"uttiker constant current (green line).
    }
    \label{FigJ}
\end{figure}

To illustrate this formula we consider an example of the reflecionless potential $V(x)= - 2/\cosh^2 x$. For this potential $T(E)=1$, hence the name.  
The Jost functions and functions $f^{(\alpha)}_q(t)$ can be easily computed and the results are presented in \ref{refLp}. 
The exact expression for the current than reads as \eqref{J}
\begin{equation}\label{J2}
    J(t) =  \int_0^\infty \frac{dq}{\pi} \rho(q)\left(q+\sin[(1+q^2)t]+2(1+q^2)\mathrm{Im}\int\limits_0^\infty \frac{dk}{\pi}  \frac{k^2}{1+k^2} \frac{e^{it(k^2-q^2)}}{(k+i0)^2-q^2}\right).
\end{equation}
We plot this expression for $\rho(E) = \theta(E_F-E)$ in figure~\ref{FigJ}(a) against the Landauer--B\"uttiker expression $J_{LB} = E_F/(2\pi)$.
Notice that even though the bound state is present in the spectrum, it produces only vanishing with time oscillations. 

To demonstrate the persistent oscillations we consider the symmetric double delta barrier potential 
\begin{equation} \label{symdoubledelta}
    V(x) = g \delta(x-d/2)+ g\delta(x+d/2).
\end{equation}
The corresponding scattering data can be computed explicitly (for the details see \ref{asff}) 
	\begin{equation}\label{Tdelta2a1}
		a_k = \frac{g^2e^{2 i k d}+(2 k+i g)^2}{4 k^2}, \quad 
			b_k = \frac{g e^{ -i d k} (g-2 i k)-g e^{ i d k} (g+2 i k)}{4 k^2}.
	\end{equation}
The bound states momenta follow from the relation $a_{i\varkappa}=0$, which if we define $u=2\varkappa/|g|$,  $D=|g|d$ can be written as 
\begin{equation}\label{bound2d1}
(u-1)^2 -e^{-u D}=0. 
\end{equation}
For the negative couplings this equation has two solutions for $D>2$ and one solution for $0\le D \le 2$. Note $a_k$ has a simple pole at $k=0$ if $D\ne 2$.
The case $D=2$ describes a situation when the bound states is just starts to appear from (disappear into) the continuous spectrum, which formally is reflected 
in $a_k$ being regular at $k=0$. Notice that same behavior is inherent for the reflectionless potentials, while for generic potentials $a_k$ has a simple pole at $k=0$.
The formula for the asymptotic current \eqref{Jtot}  is now given by 
\begin{equation}\label{Jtot2}
     J(t)\approx \int \frac{dE}{2\pi} \rho(E) T(E) +
         A_{12} \sin t(E_2-E_1),
\end{equation}
where $T(E)=|a_k|^{-2}$ is the transmission coefficient;  the energies of bound states $E_j=-\varkappa_j^2$ are defined via solutions $\varkappa_j$ of the equation \eref{bound2d1};
the amplitude $A_{12}$ follows from \eqref{Amn} and is presented explicitly in \eqref{A12d2}. 
In figure~\ref{FigJ}(b) we compare the asymptotic current \eqref{Jtot2} with the exact expression \eqref{J} 
computed numerically using $f^{(\alpha)}_q(t)$ given in \ref{asff}. We observe that the asymptotic regime establishes after few oscillations.

\section{Summary and Outlook}\label{summ}

To summarize, we have presented derivations of the Full Counting Statistics for the one-dimensional transport via an arbitrary defect from the first principles. 
The derivation in the main part is based on the effective presentation of the Green's function in the thermodynamic limit. 
The procedure of taking this limit (replacing the summation of the quantized quasimomenta to the integral) is not absolutely rigorous, so in the appendix, we have presented an exact summation over the quantized momenta with the subsequent thermodynamic limit. 
The final answer can be expressed via the Fredholm determinant whose numerical evaluation is straightforward.  

We speculate that the large-time asymptotic behavior of the obtained Fredholm determinant could be deduced after certain approximations of the kernels, which render 
the determinant to be of the sine-kernel type. 
In this form, the answer depends only on the transmission coefficient of the post-quench potential, while the correlations of the original state are present only as the energy distribution. After these approximations, the Fredholm determinant could be analyzed either by the non-linear steepest descent method for the corresponding Riemann--Hilbert 
problem or by application the effective form factors. This way we were able to recover the Levitov--Lesovik formula and its modification by logarithmic corrections in case of 
discontinuous initial distributions. 

As for the future directions, one can turn to the special quench of the \textit{perfect lead attachment} when the obtained exact kernel is an integrable one
and the Riemann--Hilbert problem appears without any approximations  (see Section~\ref{perf}). 
It would be also interesting to develop effective form factor methods to find large-time asymptotic behavior directly from the series \eqref{ff1}. 
Besides, these methods could be used to describe the situation when the Levitov--Lesovik formula is not applicable, i.e. when there are 
two or more bound states present in the spectrum of the post-quench potential and the FCS gets persistent oscillating behavior even for the constant potential bias. 
We plan to clarify how the amplitudes of these oscillations depend on the initial conditions and whether some memory effects of the pre-quench potentials are present. 

In this manuscript, we have not considered the case when there are bound states present in the pre-quench potential, but this case can be easily addressed in our formalism. 
Much more involved improvement of the formalism would be needed to tackle more general initial states (in particular, when there are some particles on the right-hand side of the system 
$\langle N_R(0)\rangle\neq 0$); to describe spinful electron and superconducting setups,  and to explore the case of the driven system i.e. when the defect depends on time (for example, for the harmonically driven conformal defect \cite{PhysRevB.103.L041405}). 

Apart from introducing the spin degrees of freedom, it would be interesting to address the multichannel scenario, which is more relevant to the theoretical description of the mesoscopic experiments. Indeed, in the typical setup, the leads are infinite in only one dimension while confinement in other dimensions creates additional channels connected with the possibility to excite transverse modes \cite{Imry1996-ft,PhysRevLett.60.848}. We expect that the corresponding Fredholm determinants for the Full Counting Statistics will contain block kernels as the transmission coefficient $T(E)$ will become a matrix. 

It is worth noting that our main assumption is based on the validity of the description of the electrons as essentially non-interacting fermions. This assumption is valid for equilibrium situations as a virtue of the Landau--Fermi theory and might be violated for non-equilibrium situations as we have here. We expect however that it remains valid as for the typical descriptions of the transport in driven nanoscale systems \cite{KOHLER2005}.
Physically, we require the existence of the quasiparticles with a lifetime sufficient for the proposed effects to be detected. In our case, this has to be larger than the frequency defined by the energy differences of two bound states.

Finally, let us mention that the experiments with ultracold atoms open a new venue to study transport in truly one-dimensional systems  \cite{PhysRevX.8.011053}. There the interactions are taken into account within the bosonization theory. We expect that the results of bosonization could be seen in the asymptotics of the corresponding Fredholm determinants (as it was for free fermions \cite{Gamayun2020}). However, the complete inclusion of interaction in the leads requires a separate investigation. There is also a full analog of the Landauer--B\"uttiker formalism for the interaction on the defect \cite{PhysRevLett.68.2512}. It would marvelous to find analogous formulas for the FCS, which is very challenging with our formalism.

\ack
We are grateful to Jakub Tworzydło and Artur Slobodeniuk for useful discussions and for careful reading of the manuscript.
The authors acknowledge support by the National Research Foundation of Ukraine grant 2020.02/0296.
Y.Z. and N.I. were partially supported by NAS of Ukraine (project No. 0122U000888).
O.G. also acknowledges support from the Polish National Agency for Academic
Exchange (NAWA) through the Grant No. PPN/ULM/2020/1/00247. O.G. is grateful to 
Galileo Galilei Institute for hospitality and support during the
scientific program on “Randomness, Integrability, and
Universality”, where part of this work was done.

\appendix

\section{Green's function calculation}
\label{appG}
In this appendix we compute the thermodynamic limit of the Green's function $G(x,y,t)$ defined as 
\begin{equation}
    G^*(x,y,t) \equiv \sum\limits_k  \frac{\chi_k(x)\chi_k(y)}{(\chi_k,\chi_k)} e^{itE_k},\qquad t\ge 0. 
\end{equation}
Here summation is taken over all solution of the spectrum condition \eref{sp23}. 
For a moment we focus on the case when bound states are absent in the spectrum. 
Using notations for $\chi_k$ in  \eref{chikk}, the norm \eref{norm} and the phase \eref{sp23}.
We present for one particular choice of the sign of the square root $\sqrt{1+({\rm Re}\,b_k)^2}$
\begin{equation}
    \frac{\chi_k(x)\chi_k(y)}{(\chi_k,\chi_k)}  = \frac{1}{2(R+\delta'(k))}{\rm Re}\, \frac{\varphi_k(x)\bar{\psi}_k(y)}{a_k} 
    - \frac{{\rm Re}\,Z_k(x,y)}{2(R+\delta'(k))\sqrt{1+({\rm Re}\,b_k)^2}}
\end{equation}
with 
\begin{equation}
   Z_k(x,y) = \frac{\psi_k(x)\psi_k(y)}{\bar a_k} + \frac{{\rm Re}\, b_k}{\bar{a}_k} \bar\varphi_{k}(x) \psi_k(y) .
\end{equation}
To evaluate the sum over $k$ we first notice that the norm \eref{norm} can be presented as a derivative of the spectrum condition \eref{sp23} 
\begin{equation}
     (\chi_k,\chi_k) =({\rm Re}\, b_k + \sqrt{1 + ({\rm Re}\, b_k)^2})\sqrt{1 + ({\rm Re}\, b_k)^2}  \frac{\partial_k [e^{2ikR+2i\delta(k)}-1]}{2i}.
\end{equation}
Further we employ the residue theorem in the following form
 \begin{equation}
     \sum_k \frac{F(k)}{\partial_k S(k)} = \frac{1}{2\pi i} \oint_\gamma  dk \frac{F(k)}{S(k)} ,
 \end{equation}
where summation is over all solutions of the equation $S(k)=0$ and the contour $\gamma$ runs around these values only and avoids any singularities of the function $F(k)$.  
This way we identically present 
\begin{equation}
    G^*(x,y,t) = \oint_\gamma \frac{dk}{2\pi} \frac{e^{itE_k}}{e^{2ikR + 2i\delta_+(k)}-1}\left(
    {\rm Re} \frac{\varphi_k(x)\bar{\psi}_k(y)}{a_k} - \frac{{\rm Re}\,Z_k(x,y)}{\sqrt{1+({\rm Re}\,b_k)^2}}
    \right)+ \left( \delta_+ \to \delta_- \right),
\end{equation}
where by $\delta_\pm$ we mean terms that are obtained by the flip of the sign  $\sqrt{1+({\rm Re}\,b_k)^2}$, specifically for the solutions of \eref{sp23}
\begin{equation}
     \frac{i {\rm Im}\,b_k+\sqrt{1+({\rm Re}\,b_k)^2}}{\bar{a}_k} \equiv e^{-2i\delta_{\rm}(k)}.
\end{equation}
The contour $\gamma$ encompasses all solutions of $e^{2ikR + 2i\delta(k)}=1$. We can present it as two contours below and above the real axes oriented in the positive and negative directions correspondingly. 
In the thermodynamic limit (with exponential accuracy) we notice that only the contour above the real line contributes, therefore we can present 
\begin{equation}
     G^*(x,y,t) =  \int\limits_{0}^{\infty} \frac{dk}{\pi} e^{itE_k}{\rm Re}\, \frac{\varphi_k(x)\bar{\psi}_k(y)}{a_k} .
\end{equation}
Here we have taken into account that upon the summation $Z_k(x,y)$ terms cancel out. 
Identically we can present
\begin{equation}
    G^*(x,y,t) =  \int\limits_{-\infty}^{\infty} \frac{dk}{2\pi} e^{itE_k}\frac{\varphi_k(x)\bar{\psi}_k(y)}{a_k} .
\end{equation}
Notice that the function that we integrate can be analytically continued to the upper half plane. 
This allows us to write the general answer in the case when bound states are present in the system as
\begin{equation}
    G^*(x,y,t) =  \int\limits_{C} \frac{dk}{2\pi} e^{itE_k}\frac{\varphi_k(x)\bar{\psi}_k(y)}{a_k} ,
\end{equation}
where the contour lies in the upper half above all positions of the bound states and connects $-\infty$ and $+\infty$. 

\section{Evaluation of $f^{(\alpha)}_q(t)$ }
\label{appF}
In this appendix we demonstrate how to rigorously evaluate $f^{(\alpha)}_q(t)$ defined in \eqref{fa2m}. We focus on $f^{(0)}_q $, as the computation for $f_q^{(1)}(t)$ goes similarly.
Namely, we are going to evaluate the thermodynamic limit of the discrete sum   
\begin{equation}
    f^{(0)}_q (t) = \sum_k \frac{(\Lambda_q,\chi_k) \chi_k(0)}{(\chi_k,\chi_k)}e^{itE_k}.
\end{equation}
The main formal problem is that the overlap $(\Lambda_q,\chi_k)$ is singular on the real line, therefore the trick with the summation introduced in \ref{appG} requires small modifications in the part choosing the integration contours. More precisely to describe the singularity we assume that without loss of generality the eigenvalues of $\Lambda_q$ and $\chi_k$ are different so the corresponding overlap could be found from 
\begin{equation}
    (k^2-q^2) (\Lambda_q,\chi_k) = \Lambda_q'(0)\chi_k(0)  - \int\limits_{-R}^0 dx \Lambda_q(x) (V_0(x) - V(x))\chi_k(x), 
\end{equation}
where we have used boundary conditions \eref{eq1} and \eref{eq2}. This way, we present
\begin{equation}
    (\Lambda_q,\chi_k) = \frac{{\rm Im} \left(e^{-i\delta(k)} \Xi^\psi_{q,k}\right)}{k^2-q^2},
\end{equation}
\begin{equation}\label{aaappp}
	\Xi^\psi_{q,k} =  \Lambda_q'(0)\psi_k(0)  - \int\limits_{-\infty}^0 dx \Lambda_q(x) (V_0(x) - V(x))\psi_k(x).
\end{equation}
Notice that here we have replaced the lower integration boundary from $-R$ to $-\infty$, which is possible due to the
finite range of the potential. 
Moreover, in this expression the dependence of the momenta $k$ and $q$ is smooth, so in particular the limit as $q\to k$ is well defined, contrary to the overall overlap, where special care has to be taken to the numerator. In particular, one can drop the quantization conditions for $k$ and consider the limit $k\to q$  
\begin{equation}
	\Xi^\psi_{q,q} =  \Lambda_q'(0)\psi_q(0)- \int\limits_{-\infty}^0 dx \Lambda_q(x) (V_0(x) - V(x))\psi_q(x). 
\end{equation}
To evaluate this expression we use the same trick as in \eqref{overlll}, \eqref{Xiqmq}, which gives 
\begin{equation}\label{ChiQQ}
    \Xi^\psi_{q,q} =  \Lambda_q'(-R)\psi_q(-R) = -\frac{q e^{iqR}}{\Phi_q(0)}\left(\bar a_q e^{iqR}- b_q e^{-iqR}\right)=- \frac{q}{\Phi_q(0)} (\bar{a}_q e^{-2i\eta(q)}- b_q).
\end{equation}
Here at the last step we have used  \eref{sp44}. For the direct proof of the result \eqref{ChiQQ} from the definition \eqref{aaappp} see \ref{pp}. 

With all these notations the function $f^{(0)}_q(t)$ can be presented as 
\begin{equation}\label{fg1}
    f^{(0)}_q(t) = \sum_k \frac{ {\rm Im} (e^{-i\delta(k)}\Xi_{q,k}){\rm Im} (e^{-i\delta(k)}\psi_k(0))}{(k^2-q^2)(\chi_k,\chi_k)}e^{itE_k}.
    % \label{fg2}
\end{equation}
We are going to evaluate the sum in  \eref{fg1} in the thermodynamic limit by presenting it as a contour integral in a way similar to \ref{appG}
\begin{equation}
    f^{(0)}_q(t) =  \oint_\gamma \frac{dk}{\pi} \frac{ {\rm Im} (e^{-i\delta(k)}\Xi_{q,k}){\rm Im}\, (e^{-i\delta(k)}\psi_k(0))}
    {(k^2-q^2)({\rm Re}\, b_k + \sqrt{1 + ({\rm Re}\, b_k)^2})\sqrt{1 + ({\rm Re}\, b_k)^2} }\frac{e^{itk^2}}{e^{2ikR+2i\delta(k)}-1}.
\end{equation}
Here contour $\gamma$ runs only around all positive solutions of the equation $e^{2ikR+2i\delta(k)}=1$ and summation over two branches  of the square root in \eref{sp23}  $\delta=\delta_{\pm}$ is assumed.
The contour $\gamma$ can be deformed into two contours above and below real line. But contrary to \ref{appG} we have to subtract
contribution from the point $k=q$, therefore we can present $f^{(0)}_q(t)$ as 
\begin{equation}\label{b10}
    f^{(0)}_q(t) =\hat{f}^{(0)}_q(t) - f^{(0,+)}_q(t) + f^{(0,-)}_q(t)  ,
\end{equation}
where 
\begin{equation}\label{B11}
   \hat{f}^{(0)}_q(t) = - i \frac{ {\rm Im} (e^{-i\delta(q)}\Xi^\psi_{q,q}){\rm Im}\, (e^{-i\delta(q)}\psi_q(0))}{q({\rm Re}\, b_q + \sqrt{1 + ({\rm Re}\, b_q)^2})\sqrt{1 + ({\rm Re}\, b_q)^2} }\frac{e^{itq^2}}{e^{2i(\delta(q)-\eta(q))}-1}
\end{equation}
and 
\begin{multline}
     f^{(0,\pm)}_q(t) = \\ 
     =\int_0^\infty\frac{dk}{\pi} \frac{ {\rm Im} (e^{-i\delta(k)}\Xi_{q,k}){\rm Im} (e^{-i\delta(k)}\psi_k(0))}{((k\pm i0)^2-q^2)({\rm Re}\, b_k + \sqrt{1 + ({\rm Re}\, b_k)^2})\sqrt{1 + ({\rm Re}\, b_k)^2} }
    \frac{e^{itk^2}}{e^{2i(k\pm i0)R+2i\delta(k)}-1}.
\end{multline}
In \eqref{B11} we have used that the point $q$ corresponds to the spectrum of the pre-quench spectrum \eref{sp44}. 
So far these transformations are exact. Further we address the large system size limit $R\to \infty$.
In this limit the last term in \eqref{b10} vanishes $ f^{(0,-)}_q\to 0$, while $f^{(0,+)}_q(t) $ can be computed identically to $G^*$ in \ref{appG}
\begin{equation}
    f^{(0,+)}_q(t)  = -\int_0^\infty\frac{dk}{\pi} 
    \frac{  {\rm Re}\left[\Xi^\varphi_{q,k}\partial_x^\alpha\bar{\psi}_k(0)a^{-1}_k \right] e^{itk^2}}{(k+ i0)^2-q^2}.
\end{equation}
To compute the residue contribution $\hat{f}^{(0)}_q(t)$ we first use  \eref{sp23} to present 
\begin{equation}
    \frac{1}{e^{2i(\delta(q)-\eta(q))}-1} = \frac{\sqrt{1+({\rm Re}\, b_q)^2}+i {\rm Im}\, b_q + a_q e^{2i\eta(q)}}{-2i{\rm Im} [a_q e^{2i\eta(q)}+b_q]},
\end{equation}
and then perform summation over all branches of the square root to obtain
\begin{equation}
       \hat{f}^{(0)}_q(t) = - \frac{1}{2q}  \frac{{\rm Re} \left[\Xi^\psi_{q,q}\partial_x^\alpha\psi_q(0)\bar{a}^{-1}_q\right]
       -(a_q e^{2i\eta(q)}-\bar{b}_q){\rm Re}\left[\Xi^\varphi_{q,q}\partial_x^\alpha\bar{\psi}_q(0)a^{-1}_q \right]}{{\rm Im} [a_q e^{2i\eta(q)}-\bar{b}_q]} e^{itq^2}.
       \end{equation}
Here  we have introduced
\begin{equation}\label{Xiphi}
    \Xi^\varphi_{q,k} \equiv a_k \Xi^\psi_{q,k}+ b_k\bar{\Xi}^\psi_{q,k} = \Lambda_q'(0)\varphi_k(0) -\int\limits_{-\infty}^0 dx \Lambda_q(x) (V_0(x) - V(x))\varphi_k(x),
    \end{equation}
which coincides with  $\Xi_{q,k}$ in the main text  (see \eqref{Xiqk}). 
The diagonal component can be obtained from \eref{ChiQQ}, 
\begin{equation}\label{tt12}
    \Xi^\varphi_{q,q} = - \frac{q}{\bar{\Phi}_q(0)},
\end{equation}
which allows us to significantly simplify expression for $\hat{f}^{(0)}_q$. Overall, for $f^{(\alpha)}_q$ we obtain the following expression 
\begin{equation}\label{Fq}
    f^{(\alpha)}_q(t) =   \frac{\partial_x^\alpha\psi_q(0)e^{itq^2}}{2i \bar{a}_q \Phi_q(0)} +  \int\limits_0^\infty \frac{dk}{\pi} {\rm Re} \left[
    \frac{\bar{\Xi}^\varphi_{q,k}\partial_x^\alpha\psi_k(0)}{\bar{a}_k}
    \right] \frac{e^{itk^2}}{(k+i0)^2-q^2}.
\end{equation}
Notice that extending the  integration over $k$ to the negative values we can also present 
\begin{equation}\label{ffqqq}
    f^{(\alpha)}_q(t) =  \int\limits_{-\infty}^\infty \frac{dk}{2\pi} 
    \frac{\Xi^\varphi_{q,k}\partial_x^\alpha\bar\psi_k(0)}{a_k}
   \frac{e^{itk^2}}{(k+i0)^2-q^2}.
\end{equation} 

Now let us discuss on how to account for the bound states. As we discussed in Section~\ref{sec2} the bound states' wave function can be understood as the Jost functions analytically continued to the upper half plane and evaluated at the purely imaginary momenta $\chi_n^{\rm bound}(x) = \varphi_{i\varkappa_n}(x)$. 
The contributions from the bound states modify  \eref{ffqqq} as follows 
\begin{equation}\label{fa1}
	f^{(\alpha)}_q(t) = \sum\limits_{n=1}^{N^{\rm b}} \frac{(\Lambda_q,\varphi_{i\varkappa_n})\partial_x^\alpha\varphi_{i\varkappa_n}(0)}{(\varphi_{i\varkappa_n},\varphi_{i\varkappa_n})}e^{-it \varkappa_n^2}+ \int\limits_{-\infty}^\infty \frac{dk}{2\pi} 
	\frac{\Xi^\varphi_{q,k}\partial_x^\alpha\bar\psi_k(0)}{a_k}
	\frac{e^{itk^2}}{(k+i0)^2-q^2}.
\end{equation}
Using the normalization  \eref{NormBound} and the relation $\varphi_{i\varkappa} = b_{\varkappa} \bar{\psi}_{i\varkappa}$, we see that we can present $f^{(\alpha)}_q$ in the following way 
\begin{equation}\label{fa2}
	f^{(\alpha)}_q(t) =  \int\limits_C \frac{dk}{2\pi} 
	\frac{\Xi^\varphi_{q,k}\partial_x^\alpha\bar\psi_k(0)}{a_k}
	\frac{e^{itk^2}}{k^2-q^2},
\end{equation}
where the contour $C$ runs from $-\infty$ to $+\infty$ and lies in the upper-half plane above all zeroes of $a_k$. 
In this form this expression coincides with \eqref{fa2m} obtained directly by going into the thermodynamic limit on the level of the Green's function.

\section{Evaluation of $\Xi^\varphi_{q,q}$}\label{pp}

In this appendix, using definition \eqref{Xiphi}
\begin{equation}\label{chi22}
	\Xi^\varphi_{q,q} \equiv \Xi_{q,q} = \Lambda_q'(0)\varphi_q(0) -\int\limits_{-\infty}^0 dx \Lambda_q(x) (V_0(x) - V(x))\varphi_k(x) ,
\end{equation}
we prove that 
\begin{equation}
	\Xi_{q,q} =- \frac{q}{\bar\Phi_q(0)},
\end{equation}
which is the statement  \eqref{tt12}. Taking into account that $\Xi_{q,-q} = \bar{\Xi}_{q,q}$ we obtain \eqref{Xiqmq}. 
Finally, the statement \eqref{ChiQQ} can be considered as a sequence of these two results and the relation $\Xi_{q,k} = a_k \Xi^\psi_{q,k}+ b_k\bar{\Xi}^\psi_{q,k}$. 

We start the proof by noticing that from the integral presentation for the Jost solutions $\Phi_q$
\begin{equation}
	\Phi_q(x) = e^{-iq x} + \int\limits^x_{-\infty} \frac{\sin(q(x-y))}{q}V_0(y) \Phi_q(y) dy,
\end{equation}
one can immediately obtain 
\begin{equation}
	\Phi_q(0) = 1 - \int\limits^0_{-\infty} \frac{\sin(qy)}{q}V_0(y) \Phi_q(y) dy,    
\end{equation}
\begin{equation}
	\Phi'_q(0) = -iq + \int\limits^0_{-\infty} \cos(qy)V_0(y) \Phi_q(y) dy.
\end{equation}
So 
\begin{equation}\label{f1}
	\Phi'_q(0) +iq \Phi_q(0) = \int\limits^0_{-\infty} e^{-iqy}V_0(y) \Phi_q(y) dy   
\end{equation}
and 
\begin{equation}\label{f2}
	\Phi'_q(0) -iq \Phi_q(0) +2iq = \int\limits^0_{-\infty} e^{iqy}V_0(y) \Phi_k(y) dy.
\end{equation}
Invoking notation for the hard-wall wave function \eref{lambda1}
\begin{equation}
	\Lambda_q(x) = {\rm Im}\frac{\Phi_q(x)}{\Phi_q(0)},
\end{equation}
we see that \eref{chi22} can be written as 
\begin{equation}\label{chi33}
\Xi_{q,q} =\frac{ 1}{2i} \left(
\frac{\Phi_q'(0)\varphi_q(0)-I_1}{\Phi_q(0)} - 
\frac{\bar\Phi_q'(0)\varphi_q(0)-\bar{I}_2}{\bar\Phi_q(0)}
\right),
\end{equation}
where 
\begin{equation}
	I_1 = \int\limits_{-\infty}^0 dx \Phi_q(x) (V_0(x) - V(x))\varphi_q(x), \qquad 
	I_{2} =  \int\limits_{-\infty}^0 dx \Phi_q(x) (V_0(x)-V(x))\bar{\varphi}_q(x) .
\end{equation}
Using integral presentation for $\varphi_q(x)$ in the first term and for $\Phi_q(x)$ in the second term we obtain
\begin{multline}
	I_{1} =  \int\limits_{-\infty}^0 dx \Phi_q(x) V_0(x)e^{-iqx}- \int\limits_{-\infty}^0 dx e^{-iqx} V(x)\varphi_q(x) \\
	+\int\limits_{-\infty}^0 dx \int\limits^x_{-\infty}dy \frac{\sin(q(x-y))}{q} \Phi_q(x) V_0(x)V(y)\varphi_q(y)\\- \int\limits_{-\infty}^0 dx \int\limits^x_{-\infty}dy \frac{\sin(q(x-y))}{q} \Phi_q(y) V_0(y)V(x)\varphi_q(x).
\end{multline}
Changing variables in the last two integrals, we arrive at
\begin{multline}
	I_{1} =  \int\limits_{-\infty}^0 dx \Phi_q(x) V_0(x)e^{-iqx}- \int\limits_{-\infty}^0 dx e^{-iqx} V(x)\varphi_q(x) \\
	+\int\limits_{-\infty}^0 dx \int\limits^0_{-\infty}dy \frac{\sin(q(x-y))}{q} \Phi_q(x) V_0(x)V(y)\varphi_q(y).
\end{multline}
Presenting sine in the exponential form and substituting right hand sides of \eref{f1} and \eref{f2} we obtain
\begin{equation}
	I_1 = \Phi_q'(0)\varphi_q(0) - \varphi_q'(0) \Phi_q(0).
\end{equation}
Similarly we can compute $I_2$
\begin{equation}
	I_2 = 2iq + \Phi_q'(0)\bar\varphi_q(0) - \bar\varphi_q'(0) \Phi_q(0).
\end{equation}
Substitution of $I_1$ and $I_2$ into \eqref{chi33} finishes the proof. 

\section{Kernels and scattering data for specific potentials}

\subsection{Single delta potential}\label{singleDelta}

In this appendix we present explicit formulas for the scattering data and the FCS for the 
quench situation that corresponds to $V(x)=g\delta(x)$, $V_0(x)=0$.

The Jost functions can easily found from the integral presernations \eref{psiint} and \eref{phiint} 
\begin{equation}
    \psi_k(x) =  e^{-ikx} - \frac{g}{k} \theta(-x) \sin (kx),
\end{equation}
\begin{equation}
    \varphi_k(x) = e^{-ikx} +  \frac{g}{k} \theta(x)  \sin (kx),
\end{equation}
where $\theta(x)$ is Heaviside step function.
The scattering data can be immediately read off from this presentation 
\begin{equation}\label{Tdelta}
    a_k = 1- \frac{g}{2ik},\qquad b_k = \frac{g}{2ik},\qquad 
    T(E)  = \frac{1}{|a_k|^2} = \frac{k^2}{k^2+g^2/4} = \frac{E}{E+g^2/4}.
\end{equation}
To describe bound states we introduce $\varkappa=|g|/2$.
If $g<0$  the bound state corresponds to the zero of $a_k$ at the momentum
$k=i\varkappa$. The corresponding wave function reads
\begin{equation}
    \varphi_{i\varkappa}(x) = e^{-\varkappa |x|}.
\end{equation}
The Jost and hard-wall wave functions corresponding to the initial potential $V_0(x)=0$ are
\begin{equation}\label{LambdaV00}
    \Phi_q(x)=e^{-iqx},\qquad \Lambda_q(x)=\mathrm{Im}\, \Phi_q(x)=-\sin qx.
\end{equation}
This leads to $\Xi_{q,k}=\Lambda_q'(0)\varphi_k(0)=-q$.
Using presentation \eqref{faRe} we obtain
\begin{equation}
    f_q^{(1)}(t)=\frac{1}{2}qe^{itq^2},
\end{equation}
\begin{equation}
    f^{(0)}_q(t)=-\frac{1}{2}\frac{qe^{itq^2}}{iq+g/2}-\theta(-g)\frac{q\varkappa e^{-it\varkappa^2}}{\varkappa^2+q^2}+qE_\varkappa(q),
\end{equation}
where
\begin{equation}\label{Ekappa}
	E_\varkappa(q) = \int\limits_0^\infty \frac{dp}{\pi} \frac{p^2 e^{itp^2}}{(p^2+\varkappa^2)((p+i0)^2-q^2)}=\frac{\varkappa \bar{h}_{\varkappa}(t)}{2(q^2+\varkappa^2)}
	- \frac{iq h_q(t)}{2(q^2+\varkappa^2)}.
\end{equation}
and
\begin{equation}
	h_q(t) = e^{itq^2}\left[1- {\rm Erf} \left(qe^{i\pi/4}\sqrt{t}\right)\right].
\end{equation}
The FCS can be written as 
\begin{equation}
	\mathcal{F} (\lambda,t) = \det \left(1 + \frac{e^\lambda-1}{\pi}\rho(q)X_0(q,q') +\frac{e^\lambda-1}{\pi}\rho(q)X_1(q,q')\right),
\end{equation}
where 
\begin{equation}
	X_0(q,q') =qq' \left(\frac{q}{\varkappa^2+q^2}+\frac{q'}{\varkappa^2+q'^2}\right) \frac{\sin\left[t(q^2-q'^2)/2\right]}{q^2-q'^2},
\end{equation}
\begin{multline}
	X_1(q,q') = 
	-2q q'\mathrm{Im}\left(e^{-it(q^2+q'^2)/2}\frac{e(q)-e(q')}{q^2-q'^2}\right) \\
	+\frac{q q'}{(\varkappa^2+q^2)(\varkappa^2+q'^2)}\left\{\varkappa\,\mathrm{Re}
	\left(e^{it(q^2+q'^2)/2}h_{\varkappa}(t)\right)\right.\\
	\left. 
	-\frac{g}{2}\cos\left[t(q^2-q'^2)/2\right]-2\theta(-g)\varkappa\cos \left[t(q^2+q'^2+2\varkappa^2)/2\right]\right\},
\end{multline}
and 
\begin{equation}
    e(q)=\frac{q h_q(t)}{2(q^2+\varkappa^2)}.
\end{equation}
In the notations of \eref{Kqqp}  $K_0 = \rho (e^\lambda-1)X_0$ and $\delta K = \rho (e^\lambda-1)X_1$. 

The propagation emerging from a step initial distribution formally corresponds to $V_0(x)=0$ for  $x<0$ and $V(x)=0$.
The corresponding FCS can be obtained from the above formulas by simply sending $g\to 0$. The corresponding kernels simplify as follows
\begin{equation}
	X_0(q,q') =(q+q') \frac{\sin\left[t(q^2-q'^2)/2\right]}{q^2-q'^2},
\end{equation}
% \begin{equation}
% 	X_1(q,q') = 
% 	-2q q'\mathrm{Im}\left(e^{-it(q^2+q'^2)/2}\frac{e(q)-e(q')}{q^2-q'^2}\right),
% 	\qquad
% 	 e(q)=\frac{h_q(t)}{2q}.
% \end{equation}
\begin{equation}
	X_1(q,q') = 
	-\mathrm{Im}\left(e^{-it(q^2+q'^2)/2}\frac{q'h_q(t)-qh_{q'}(t)}{q^2-q'^2}\right).
\end{equation}

\subsection{Reflectionless potential}\label{refLp}

In this appendix we consider an example of perfect lead attachment, i.e. $V_0(x)=V(x)$, $x<0$, for the reflectionless potential 
\begin{equation}\label{Vcosh}
V(x)=-\frac{2}{\cosh^2 x}.
\end{equation}
The corresponding Jost solutions are 
\begin{equation}
    \psi_k(x)= e^{-ikx}\left(1+\frac{2i}{k-i}\frac{1}{e^{2x}+1}\right),
\end{equation}
\begin{equation}
    \varphi_k(x)=\bar\psi_k(-x)=e^{-ikx}\left(1-\frac{2i}{k+i}\frac{1}{e^{-2x}+1}\right)=\frac{k-i}{k+i} \psi_k(x),
\end{equation}
which lead to the following scattering data
\begin{equation}
    a_k=\frac{k-i}{k+i} , \qquad b_k=0.
\end{equation}
This potential has one bound state corresponding to the zero of $a_k$ at $k=i$:
\begin{equation}
   \chi_1^{\rm b}(x) = \varphi_{k=i}(x)=\frac{1}{2\cosh x}.
\end{equation}
The hard-wall wave functions defined in \eref{lambda1} are given by
\begin{equation}
    \Lambda_q(x)=-\frac{q \sin qx +\tanh x\cos qx}{q}.
\end{equation}
Therefore $\Xi_{q,k}$ in \eref{Xiqk} becomes
\begin{equation}
    \Xi_{q,k}= \Lambda_q'(0)\varphi_k(0) =-\frac{1+q^2}{q} \cdot \frac{k}{k+i}.
\end{equation}
Using definitions \eref{faRe}  we arrive at 
\begin{equation}
    f^{(1)}_q(t)=-\frac{1+q^2}{2q}e^{itq^2},
\end{equation}
\begin{equation}\label{GqCosh}
    f^{(0)}_q(t) =  \frac{e^{-it }}{2q}
    +\frac{e^{itq^2}}{2i} 
    -\frac{1+q^2}{q} E_1(q)
\end{equation}
where $E_\varkappa$ is defined in \eqref{Ekappa}. 
Substituting these expression into \eqref{J} we arrive at \eqref{J2}.

\subsection{Double delta barrier} \label{asff} 

The double delta barrier potential is given by 
	\begin{equation}\label{V2delta}
		V(x) = g_1 \delta(x-d_1)+g_2\delta(x-d_2),  
	\end{equation}
where we assume that $d_2>0>d_1$. The Jost solutions for this potential can be found via the integral presentations \eref{psiint} and \eref{phiint} 
	\begin{equation}
		\psi_k(x) = e^{-ik x} - \theta(d_1-x)\frac{\sin(k(x-d_1))}{k}g_1 \psi_k(d_1) - \theta(d_2-x)\frac{\sin(k(x-d_2))}{k}g_2 \psi_k(d_2),
	\end{equation}
	\begin{equation}
		\varphi_k(x) = e^{-ik x} 	+\theta(x-d_1)\frac{\sin(k(x-d_1))}{k}g_1 \varphi_k(d_1) +\theta(x-d_2)\frac{\sin(k(x-d_2))}{k}g_2 \varphi_k(d_2),
	\end{equation}
where 
	\begin{equation}
	\psi_k(d_1) = e^{-ikd_1 } \left(1 + \frac{g_2}{2ik} \right)  - \frac{g_2}{2ik}e^{ik(d_1-2d_2)}, \qquad 
		\psi_k(d_2)= e^{-ikd_2},
	\end{equation}
	\begin{equation}
		\varphi_k(d_1) = e^{-ikd_1}, \qquad 
		\varphi_k(d_2) =  e^{-ikd_2 } \left(1 -\frac{g_1}{2ik} \right)  +\frac{g_1}{2ik}e^{ik(d_2-2d_1)}.
	\end{equation}
The scattering data follows from \eref{transfer} 
	\begin{equation}\label{Tdelta2a}
		a_k = \frac{g_1 g_2 e^{-2 i k (d_1-d_2)}+(2 k+i g_1) (2 k+i g_2)}{4 k^2},
	\end{equation}
	\begin{equation}\label{Tdelta2b}
		b_k = \frac{g_2 e^{-2 i d_2 k} (g_1-2 i k)-g_1 e^{-2 i d_1 k} (g_2+2 i k)}{4 k^2}.
	\end{equation}
If we were interested only in the scattering data we could easily found them using results of \ref{singleDelta}.
Indeed, for any potential that can be presented as a  disjoint sum i.e. $V(x)=V_1(x)+V_2(x)$ with
$V_1(x)=0$ for $x>x_1$ and $V_2(x)=0$ for $x<x_2$, where $x_1<x_2$, the transfer matrix reads 
	\begin{equation}
		\mathcal{T}=\mathcal{T}_1\mathcal{T}_2,
	\end{equation} 
where $\mathcal{T}_j$ is the transfer matrix for $V_j$. This statement follows immediately from the 
relation of $\mathcal{T}_j$ to the corresponding Jost solutions $\psi_j$ and $\varphi_j$, namely, 
\begin{equation}
	\begin{pmatrix}
		\varphi_1\\
		\bar{\varphi}_1
	\end{pmatrix}=	\mathcal{T}_1\begin{pmatrix}
	\psi_1\\
	\bar{\psi}_1
\end{pmatrix}=\mathcal{T}_1 \begin{pmatrix}
	\varphi_2\\
	\bar{\varphi}_2
\end{pmatrix}=\mathcal{T}_1\mathcal{T}_2\begin{pmatrix}
\psi_2\\
\bar{\psi}_2
\end{pmatrix}.
\end{equation}
Further, taking into account that the transfer matrix $\tilde {\mathcal{T}}$ for the shifted potential $\tilde V(x)=V(x-d)$ is related to $\mathcal{T}$ 
by conjugation with a diagonal matrix
\begin{equation}
\tilde{\mathcal{T}}=\mathcal{T}(d)=\begin{pmatrix}
		a_k & b_k e^{-2ikd}\\
		\bar{b}_k e^{2ikd} & \bar{a}_k 
	\end{pmatrix},
\end{equation}
the scattering data \eref{Tdelta2a} and \eref{Tdelta2b} for the potential \eref{V2delta} is recovered from
\begin{equation}
	\mathcal{T}=\mathcal{T}_{g_1}(d_1)\mathcal{T}_{g_2}(d_2), \qquad  
	\mathcal{T}_g(0)=\begin{pmatrix}
		1-\frac{g}{2ik} & \frac{g}{2ik}\\
		-\frac{g}{2ik} & 1+\frac{g}{2ik}
	\end{pmatrix},
\end{equation}
where for $\mathcal{T}_g(0)$ we used \eref{Tdelta}. 
The bound states correspond to zeroes of $a_k$ in the upper half plane of $k$. 
For negative coupling constants $g_1$ and $g_2$ we have two bound states if 
\begin{equation}\label{bound2d}
	d_{2}-d_1 > \frac{1}{|g_1|}+\frac{1}{|g_2|},
\end{equation}
and one otherwise.

The symmetric potential corresponds to   
$g_1=g_2=g$, $d_2=-d_1=d/2$. We introduce notations 
\begin{equation}
k=i\varkappa, \qquad
	u=2\varkappa/|g|>0, \qquad D=|g|d,
\end{equation}
so that the quantity $\varkappa$ describes the ``momentum" of the bound state.  
The condition \eqref{bound2d} now reads $D>2$ (see also discussion around equation \eqref{bound2d1}).  
The current and the kernel in this case are obtained by the numerical integration of the corresponding expressions constructed via $f^{(\alpha)}_q(t)$ in \eref{faRe}. 
For the case when $V_0(x)=0$ we have \eref{LambdaV00}.
Hence 
\begin{equation}\label{Xi2delta}
	\Xi_{qk}=\Lambda'_q(0)\varphi_k(0)+\int_{-\infty}^{0}dx \Lambda_q(x)V(x)\varphi_k(x)=-q-g  e^{ikd/2} \left(\frac{q}{k}\sin \frac{kd}{2}-\sin \frac{qd}{2}\right),
\end{equation}
and 
\begin{equation}
	f^{(1)}_q(t) =  
	 B_{2,q}^{(1)} e^{-it\varkappa_2^2}
		+F^{(1)}_q e^{itq^2} + I^{(1)}_q(t),
\end{equation}
\begin{equation}
	f^{(0)}_q(t) =  
	 B_{1,q}^{(0)} e^{-it\varkappa_1^2}
		+F^{(0)}_q e^{itq^2} + I^{(0)}_q(t),
\end{equation}
\begin{equation}
    B_{n,q}^{(\alpha)}=
    \frac{i \Xi_{q,i\varkappa_n}\partial_x^\alpha\bar \psi_{i\varkappa_n}(0)}
    {a'_{i\varkappa_n}(\varkappa_n^2+q^2)},
    \qquad
    F_q^{(\alpha)}=-i \frac{\partial_x^\alpha\psi_{q}(0)}{2a_{-q}},
\end{equation}
\begin{equation}
	a'_{i\varkappa_j}=\left.\frac{da}{dk}\right|_{k=i\varkappa_j}=-\frac{2i}{|g|}\frac{(u_j-1)(D(u_j-1)+2)}{u_j^2},
\end{equation}
\begin{equation}
	\bar\psi_{i\varkappa_1}(0)=2-2/u_1, \qquad  \bar\psi_{i\varkappa_2}(0)=\partial_x \bar\psi_{i\varkappa_1}(0)=0, \qquad 
	\partial_x \bar\psi_{i\varkappa_2}(0)=(1-u_2)|g|.
\end{equation}
The integrals are given by
\begin{equation}
	I^{(\alpha)}_q(t)=\int\limits_{0}^{\infty} \frac{dk}{\pi} 
	\Omega^{(\alpha)}_{q,k}
	\frac{e^{itk^2}}{(k+i0)^2-q^2},
\end{equation}
with
\begin{equation}
	\Omega^{(0)}_{q,k}=\frac{2k^2(-q+g\cos kd/2\sin qd/2)}{g^2+2k^2+g^2\cos kd-2gk \sin kd},
\end{equation}
\begin{equation}
	\Omega^{(1)}_{q,k}=-\frac{2gk^3\sin kd/2\sin qd/2}{g^2+2k^2-g^2\cos kd+2gk \sin kd}.
\end{equation}
The asymptotic behavior of the integrals $I^{(\alpha)}_q(t)$ at large $t$ is governed by expansions of the integrands at $k=0$
\begin{equation}
	\Omega^{(0)}_{q,k}=\frac{k^2}{g^2}(-q+g\sin qd/2)+O(k^4), \quad \Omega^{(1)}_{q,k}=-\frac{2k^2gd\sin qd/2}{(2+gd)^2} + O(k^4).
\end{equation}
The formula for $\Omega^{(1)}_{q,k}$ is valid for $D=-gd\ne 2$. The asymptotic   behavior of $\Omega^{(1)}_{q,k}$ for  $D=2$ and small $k$ is
\begin{equation}
	\Omega^{(1)}_{q,k}=\frac{4}{d^2}\sin \frac{qd}{2}+\frac{k^2}{18}\sin \frac{qd}{2}+O(k^4).
\end{equation} 
Therefore, the integrals have the following decaying behavior for large $t$
\begin{equation}
	I^{(\alpha)}_q(t)\sim t^{-\frac{3}{2}}  \quad \mathrm{for} \quad D\ne 2,\quad  \mathrm{and} \quad I^{(\alpha)}_q(t)\sim t^{-\frac{3}{2}+\alpha} \quad \mathrm{for} \quad D=2.
\end{equation}
This demonstrates that they do not affect the leading contribution in the asymptotic current  \eref{Jtot}. If the potential has two bound states than there is an oscillatory part of the current with the amplitude of oscillations
given by \eref{Amn}
\begin{equation}\label{A12d2}
A_{12} =-\frac{4}{\pi} \int_0^\infty dq  \rho(q)  
B^{(1)}_{2,q} B^{(0)}_{1,q}.
\end{equation}
Finally,  the leading contribution to the current for large $t$ consists of
constant  Landauer--B\"uttiker current and an oscillating current (if there are two bound states).

\section*{References}

%\nocite{*}

\bibliography{lb}

\end{document}